\documentclass[12pt,preprint]{aastex}

\usepackage{soul}
\usepackage{ulem}
\newcommand{\ha}{H$\alpha$}

\newcommand{\rosat}{{\sl ROSAT}}

\newcommand{\galex}{{\sl GALEX}}

\slugcomment{Accepted to ApJ}

\shorttitle{Young M dwarfs within 25 pc II}
\shortauthors{Shkolnik et al.}

\begin{document}


\title{Identifying the young low-mass stars within 25 pc. II. Distances, kinematics \& group membership\altaffilmark{1}\\}


\author{Evgenya~L.~Shkolnik}
\affil{Lowell Observatory, 1400 W. Mars Hill Road, Flagstaff, AZ 86001}
\email{shkolnik@lowell.edu}

\author{Guillem Anglada-Escud\'e}
\affil{Universit\"at G\"ottingen, Institut f\"ur Astrophysik, Friedrich-Hund-Platz 1, 37077 G\"ottingen, Germany}
%\email{guillem.anglada@gmail.com}

\author{Michael C. Liu, Brendan P. Bowler}
\affil{Institute for Astronomy, University of Hawaii at Manoa\\ 2680 Woodlawn Drive, Honolulu, HI 96822}
%\email{mliu@ifa.hawaii.edu}

\author{Alycia J. Weinberger, Alan P. Boss}
\affil{Department of Terrestrial Magnetism, Carnegie Institution of Washington, 5241 Broad Branch Road, NW, Washington, DC 20015}

%\author{Brendan P. Bowler}
%\affil{Institute for Astronomy, University of Hawaii at Manoa\\ 2680 Woodlawn Drive, Honolulu, HI 96822}

\author{I. Neill Reid}
\affil{Space Telescope Science Institute, Baltimore, MD 21218}
%\email{inr@stsci.edu}

\and

\author{Motohide Tamura}
\affil{National Astronomical Observatory of Japan, Tokyo, Japan}

\altaffiltext{1}{Based on
observations collected at the W. M. Keck Observatory, the Canada-France-Hawaii Telescope, the du Pont Telescope at Las Campanas Observatory, and the Subaru Telescope.  The Keck Observatory is operated as a scientific partnership between the California Institute of
Technology, the University of California, and NASA, and was made possible by the generous
financial support of the W. M. Keck Foundation. The CFHT is operated by the National Research Council of Canada,
the Centre National de la Recherche Scientifique of France, and the University of Hawaii.}

\begin{abstract}

We have conducted a kinematic study of 165 young M dwarfs with ages of $\lesssim$300 Myr. Our sample is composed of stars and brown dwarfs with spectral types ranging from K7 to L0, detected by \rosat\ and with photometric distances of $\lesssim$25 pc assuming the stars are single and on the main-sequence. 
In order to find stars kinematically linked to known young moving groups (YMGs), we measured radial velocities for the complete sample with Keck and CFHT optical spectroscopy and trigonometric parallaxes for 75 of the M dwarfs 
with the CAPSCam instrument on the du Pont 2.5-m Telescope.  Due to their youthful overluminosity and unresolved binarity, the original photometric distances for our sample underestimated the distances by 70\% on average, excluding two extremely young ($\lesssim$3 Myr) objects
found to have distances beyond a few hundred parsecs. We searched for kinematic matches to 14 reported YMGs and identified
9 new members of the AB~Dor YMG and 
2 of the Ursa Majoris group. Additional possible candidates include
6  Castor,
4 Ursa Majoris,
2 AB Dor members,
and 1 member each of the Her-Lyr and $\beta$~Pic groups.
Our sample also contains 27 young low-mass stars and 4 brown dwarfs with ages $\lesssim$150 Myr which are not associated with any known YMG.
We identified an additional 15 stars which are kinematic matches to one of the YMGs, but the ages from spectroscopic diagnostics and/or the positions on the sky do not match. These warn against grouping stars together based only on kinematics and that a confluence of evidence is required to claim that a group of stars originated from the same star-forming event.

\end{abstract}


\section{Introduction}\label{intro}

Planet-formation studies require a range of stellar host mass and age in order to test models of planet and disk evolution. 
The vast majority of confirmed exoplanets are in systems greater than 1 Gyr old, and efforts to find the youngest planets, those just forming in their disks around T Tauri stars (1 -- 3 Myr old),  are just beginning \citep{prat08,croc11,krau11}. Young stars with ages spanning 10 to 100 Myr fill a particularly interesting, and relatively unexplored, gap as this time scale coincides with the end of giant planet formation (e.g.~\citealt{boss11}) and the onset of active terrestrial planet formation. 

From an observational perspective, this time range is critical for direct imaging searches of planets and disks (e.g.~\citealt{maro08,liu10}).  Planets cool and fade dramatically during this period, e.g.~the luminosity of a 5-M$_{Jup}$ planet drops by 2 -- 3 orders of magnitude in this time span and its effective temperature decreases from 1300 K to 400 K (\citealt{bara03}). Also in this time period, the fraction of debris disks around AFGK main-sequence stars decreases significantly (e.g.~\citealt{riek05,hill08}). 
These attributes have made young, nearby stars prime targets at direct imaging searches for exoplanets and circumstellar disks, as supported by the imaged planets around the 10 -- 30-Myr stars   $\beta$ Pic and HR 8799 \citep{lagr10,maro08}. 

M dwarfs offer an additional observational advantage. Their intrinsic faintness provides a more favorable contrast in star-planet flux, aiding the detection of faint planetary-mass companions. M~dwarfs also dominate the stellar mass function by number$\colon$ roughly 3 out of 4 stars in a volume-limited sample of the solar neighborhood are M~dwarfs (\citealt{reid95,boch10}). So in principle low-mass stars could represent the most common and nearest hosts of planetary systems, providing a  much larger population of targets for direct imaging of exoplanets than has been studied to date.

Searches for young low-mass stars have been carried out using X-ray (e.g.~\citealt{jeff95,webb99,mont01,shko09b}, SLR09 hereafter) and UV (\citealt{shko11,rodr11}) surveys to identify strong coronal and chromospheric emission.  SLR09 and \cite{shko11} estimated that M dwarfs with fractional X-ray and UV luminosities of $log(F_X/F_J) > -2.5$ and $log(F_{NUV}/F_J) > -4$ are younger than $\sim$300 Myr, based on comparisons with Pleiades (120~Myr) and Hyades (625 Myr) members.  However, determining a more accurate age for an individual star is more difficult. 

Over the past decade, many dispersed young stars have been kinematically linked to coeval
moving groups (e.g., \citealt{zuck04,torr08}, and references therein), for which more accurate ages are available through comparison of the bulk group properties with stellar isochrones and lithium-depletion models (e.g.~\citealt{sode10}). 
Thus, pinning an X-ray- or UV- bright M dwarf to one of the known young moving groups (YMGs) provides a more accurate means to determine its age.

The current census of YMG members is mostly limited to AFGK-type stars due to the past reliance on optical catalogs for distances and proper motions, e.g., the Hipparcos and Tycho catalogs \citep{perr97}, which exclude most of the fainter nearby M dwarfs. If the initial mass function of YMGs follows the field, we can expect that there are many unidentified low-mass members of the known YMGs, e.g.~$\approx$60 M dwarfs missing from the $\beta$~Pic YMG and $\approx$20 mid-Ms from the TW Hydrae Association \citep{shko11}. Recent searches for these low-mass members using large photometric and proper motion catalogs have found many candidates awaiting confirmation with follow-up parallaxes and radial velocities (e.g., \citealt{lepi09,schl12}).

This paper presents the kinematic analysis of the 165 young M dwarfs first described in SLR09,\footnote{There are 10 additional stars in this paper compared to SLR09, including 8 common proper-motion companions found during the CAPSCam astrometric observations and 2  young M dwarfs observed in support of the Gemini NICI Planet-Finding Campaign \citep{liu10}.} including radial velocities (RVs) for all the targets, trigonometric parallaxes for half the sample, and three-dimensional space motions with the goal of providing more accurate ages by linking  these stars to known YMGs.

\section{The Sample}\label{sample}

Our original sample of nearby low-mass stars was drawn from the NStars census \citep{reid03a,reid04} and the Moving-M sample (\citealt{reid07a}), which contained  $\approx$1100 M dwarfs.  
Distances were mainly available through spectrophotometric relations as only 11\% had parallaxes. We limited the sample to a photometric distance of 25 pc assuming the stars are single and on the main sequence (e.g., \citealt{reid02c,cruz03}).
This sample was cross-matched with the \rosat\ All-Sky Survey Bright Source Catalog and Faint Source Catalog \citep{voge99,voge00} as X-ray activity is a powerful indicator of youth 
(e.g.~\citealt{prei05}). Target stars were chosen to have high X-ray emission ($log(F_X/F_J) > -2.5$) comparable to or greater than that of Pleiades members (120 Myr, \citealt{stau98}). 
A more detailed description of the sample selection process can be found in SLR09, where we used high-resolution optical spectra to determine spectral types (SpT) and eliminate spectroscopic binaries (SB; \citealt{shko10}).  We used spectroscopic age diagnostics such as low surface gravity, \ha\ emission and lithium absorption to provide upper and lower age limits for each of the targets.

\section{The Radial Velocities}\label{rv}

We acquired high-resolution \'echelle spectra of 155 (of the 165) young M dwarfs in the SLR09 sample with the High Resolution \'Echelle Spectrometer (HIRES; Vogt et al.~1994) on the Keck I 10-m telescope and  with the \'Echelle SpectroPolarimetric Device for
the Observation of Stars (ESPaDOnS; Donati et al.~2006) on the Canada-France-Hawaii 3.6-m telescope, both located on the summit of Mauna Kea. These spectra were first described in SLR09.

We used the 0.861$\arcsec$ slit with HIRES to give a spectral resolution of $\lambda$/$\Delta\lambda$$\approx$58,000. The detector consists of a mosaic of three 2048 $\times$ 4096 15-$\micron$ pixel CCDs, corresponding to a blue, green and red chip spanning 4900 -- 9300 \AA. To maximize the throughput near the peak of a M dwarf spectral energy distribution, we used the GG475 filter with the red cross-disperser. 

ESPaDOnS is fiber fed from the Cassegrain to Coud\'e focus where the fiber image is projected onto a Bowen-Walraven
slicer at the spectrograph entrance. With a 2048$\times$4608-pixel CCD detector, ESPaDOnS'
`star+sky' mode records the full spectrum over 40 grating orders covering 3700 to 10400 \AA\/ at a spectral
resolution of $\lambda$/$\Delta\lambda$$\approx$68,000. The data were reduced using {\it Libre-ESpRIT}, a fully automated reduction package
provided for the instrument and described in detail by Donati et al.~(1997, 2007).

Each stellar exposure was bias-subtracted and flat-fielded for pixel-to-pixel sensitivity variations. After optimal
extraction, the 1-D spectra were wavelength calibrated with a Th-Ar arc. Finally
the spectra were divided by a flat-field response curve and corrected to the heliocentric velocity.
The final spectra were of moderate S/N, reaching  20 -- 50 per pixel at 7000 \AA. Each night, spectra were
also taken of an A0V standard star for telluric correction and an early-, mid-, and/or late-M RV standard.\footnote{See Table~3 of SLR09 for the list of RV standards used.}

We cross-correlated each of 7 orders between 7000 and 9000 \AA\/ of each stellar spectrum with an RV standard of similar spectral type using the  {\it fxcor}
routine (Fitzpatrick 1993) in IRAF.\footnote{IRAF (Image Reduction and Analysis Facility) is distributed by
the National Optical Astronomy Observatories, which is operated by the Association of Universities for Research in Astronomy, Inc.~(AURA) under cooperative agreement with the National Science Foundation.}
We measure the RVs from the gaussian peaks fitted to the cross-correlation function, taking the average and RMS of all orders as the final measurements. The RMS was typically less than 1 km~s$^{-1}$ for both instruments. (Multi-lined spectroscopic binaries were removed from the sample and reported in \citealt{shko10}.)
 The RVs, SpTs and other relevant information for each target are listed in Table~\ref{table_targets}.

\section{Astrometric Observations}

We measured parallaxes for the 75 targets (45\% of our sample) observable with the CAPSCam
instrument mounted on the 2.5-m du Pont Telescope at Las Campanas Observatory.
An additional 8 stars had published parallaxes. CAPScam was specifically
designed for high-accuracy astrometry of M dwarfs in a bandpass covering 8000 \AA\ --
9000 \AA. Its pixel size is 0.1956\arcsec\ with a field of view of
6.63\arcmin\ $\times$ 6.63\arcmin. Since the stars in this program are relatively bright (I = 9 -- 12 mag),  we used the guide window to obtain short integrations of 0.5 to 1 sec. while exposing the full field of view for 30 to 60 sec. We used this instrument to obtain between 3 and 6
astrometric epochs on each of 75 targets, with a minimum time baseline of 1.5
years.  The astrometric analysis was performed with the ATPa software developed
in the framework of the Carnegie Astrometric Planet Search Program (CAPS). A
more detailed description of the data analysis techniques can be found in
\citet{boss09} and \cite{angl11}.

In order to measure the parallax, one has to measure the motion of
a star with respect to background sources. The motion on a local
tangent plane to the sky is modeled using the 5 astrometric observables:
 R.A.~and Dec.~of the initial positions ($\alpha_0$ and
$\delta_0$), proper motion in R.A. and Dec. ($\mu_{\alpha}$ and $\mu_{\delta}$)
and parallax ($\pi$).  The local tangent plane formalism is outlined
in the Hipparcos catalog reference guide \citep{perr97} where the motion of a star
is given by

\begin{eqnarray}
X &=& \alpha_0 + \mu_\alpha \left(t - t_0\right) - \pi p_\alpha\left[t\right]\\
Y &=& \delta_0 + \mu_\delta \left(t - t_0\right) - \pi p_\delta\left[t\right]
\end{eqnarray}

\noindent where $p_\alpha$ and $p_\delta$ are the so-called parallax factors
which are the projected motion of a star with $\pi=1$ on the tangent plane of
the sky defined by R.A. and Dec. positions, $X$ and $Y$, respectively. The
reference epoch $t_0$ implicitly defines the time against which all the quantities are
referenced, and in our case, we use the last epoch of the observations.

In order to measure the motion of a star with respect to the background objects,
the field distortions of each image first need to be calibrated. To do this, a
subset of stable reference stars are used to fit a 2-d polynomial to each image.
Given the measured local plane position $x^{obs}_i$ and $y^{obs}_i$ of each
star on the detector in pixel coordinates, the calibrated position $X_i$ and $Y_i$ should be

\begin{eqnarray}
X &=& a_0 + a_x x + a_y y + a_{xx} x^2 + a_{yy} y^2 + a_{xy} xy + 
\ldots \label{eq:xcalib}\\
Y &=& b_0 + b_x x + b_y y + b_{xx} x^2 + b_{yy} y^2 + b_{xy} xy + 
\ldots \label{eq:ycalib}
\end{eqnarray}

\noindent where the $a$ and $b$ coefficients need to be determined for each
image. For this program, we used a second order distortion (terms up to $x^2$,
$y^2$ and $xy$) which required fitting 6 coefficients for each axis. This means
that, at the very least, 6 reference stars are required for each field on each
image. We used $\geq$10 reference stars in each field. Note that the absolute
field geometry and the positions of the reference stars are not known a priori
with enough accuracy. Because of this, the fitting process is iterative. In a
first iteration, the nominal positions $X$ and $Y$ of all the stars are taken
from a high quality image and all the other images are matched to it. Then an
astrometric solution fit is obtained and a refined catalog of source
positions, proper motions and parallaxes is generated. Those stars showing
smaller residuals are used as the references in the following iterations. As a
result, the astrometric solution for the target, as well as for all the other
stars in the field, is obtained. This process is iterated between 3 to 5 times
until convergence is reached (i.e.~when the RMS improves less than 0.1 mas from
one iteration to the next). After the fitting procedure, a typical target star
has an RMS between 0.6 and 1.5 mas/epoch. 

All the processing is done using the
ATPa software specifically developed for CAPSCam. The software controls the
end-to-end processing, which consists of centroid and source extraction,
crossmatching, distortion, and calibration, producing final astrometric products.  These
products are source catalogs with the astrometric solutions and the motion of
each star in the field. From measurements on many other fields within the CAPS
project \citep{boss09}, the absolute orientation of CAPScam is known to be
stable down to 10$\arcmin$ RMS in position angle (X is East-West, Y is
North-South). In the small angle approximation, this angle, 0.0029
radians, is the maximum relative systematic error induced on any measured
small offset. This is, all measured displacements (e.g.~due to parallax 
and proper motion) contain a relative uncertainty of 0.3\% due the unknown
absolute field rotation. The plate scale has also been measured to be 
constant at 0.5\% precision so both effects added in quadrature amounts
to 0.6\%. Given that this is significantly smaller than our aimed precision of the parallaxes (5\%), nominal orientation and scale 
of CAPScam were assumed for all the fields and the associated uncertainties 
were not included in the error budget for the distance determinations.

In addition to the astrometric parameters, the pipeline also provides a Monte
Carlo estimate of the uncertainties. This is important when the number of
independent measurements (twice the number of epochs because each measurement is
two dimensional) is similar to the number of unknowns (i.e.~$\sim$5 astrometric
parameters). The formal uncertainties derived from the equivalent least squares
problem are typically too optimistic in this situation. Instead, we simulated
1000 datasets evaluated at the same epochs of observation and added Gaussian
noise with an empirically determined uncertainty on each direction. The epoch
uncertainty is estimated by computing the standard deviation of the residuals of
the reference stars over all the epochs (not individual images). From other
CAPSCam programs \citep[e.g.][]{boss09}, we know that the typical uncertainty
for 20 minutes on sky integration is usually better than 1.5~mas. When the
standard deviation obtained from the references is lower than this typical
precision, we conservatively assume a single epoch uncertainty of 1.5 mas. This
procedure allowed us to obtain realistic uncertainties and account for parameter
correlation (e.g.~proper motion and parallax) robustly. Note that this is an
approximate but conservative approach because it assumes the same empirically
determined uncertainty on all the epochs, even if there is a very poor quality
one contaminating the measure of the uncertainty. Given the low number of
observations compared to the free parameters and because astrometric
measurements tend to be dominated by epoch-to-epoch systematic effects, we
consider that this Monte Carlo approach provides the most secure and reliable
estimation of the uncertainties in the astrometric observables.

In several instances, the RMS of the residuals of the target star is
significantly larger than expected, e.g.$\sim$ 4 mas RMS, when a typical star in
the same field has 1.6 mas RMS. We found this occurs when the target appears to
be a barely resolved visual binary (VB) with a separation smaller than
1.5${\arcsec}$. In these cases, we assigned a more conservative 5-mas uncertainty
to the parallax measurement. There are a few stars with excess scatter but no
evidence for binarity. All the barely-resolved binaries and high-RMS stars are
flagged in Table \ref{tab:astrodata}.

The last important issue  is that parallax measurements
are made relative to the reference stars. The reference stars are not at an
infinite distance and thus have a finite parallactic motion as well. Because of
the matching process in the geometric calibration step (Equations
\ref{eq:xcalib} and \ref{eq:ycalib}), the average parallactic motion of the
references will be absorbed so there will be a zero-point ambiguity in the
measurement of the distance to the target. Correcting for the zero-point
parallax requires making an educated guess of the distances to the reference
stars, which may be complicated by the fact that good homogeneous photometry is
not available in all cases. Instead, we estimated the typical zero-point
correction to a few fields and applied it to the rest of the dataset. 
This procedure consists of using the photometry in the NOMAD catalog ($B, J, H$ and $K$), estimating the photometric distance to a few reference stars, and computing the
average difference between the measured parallax and the photometric one. Further
details on the zero-point correction computation can be found in \citet{angl11}. In the
fields we tested, the zero-point correction was found to be smaller than 1 mas
in all cases, with an average value of approximately +0.5 mas. Given that the precision
in the relative parallax is never better than 1 mas, we add in quadrature the
0.5 mas obtained from the test fields to their statistical uncertainties of the measured parallaxes. Note that our typical
target lies within 30 pc ($\pi$=33 mas), meaning that a 0.5-mas uncertainty
represents only a 2\% systematic error in the actual distance (compared to the
 1.5 mas or 5\% contribution derived from typical random errors in the
measured relative parallaxes.) 

\begin{figure}
\epsscale{1.0}
\plotone{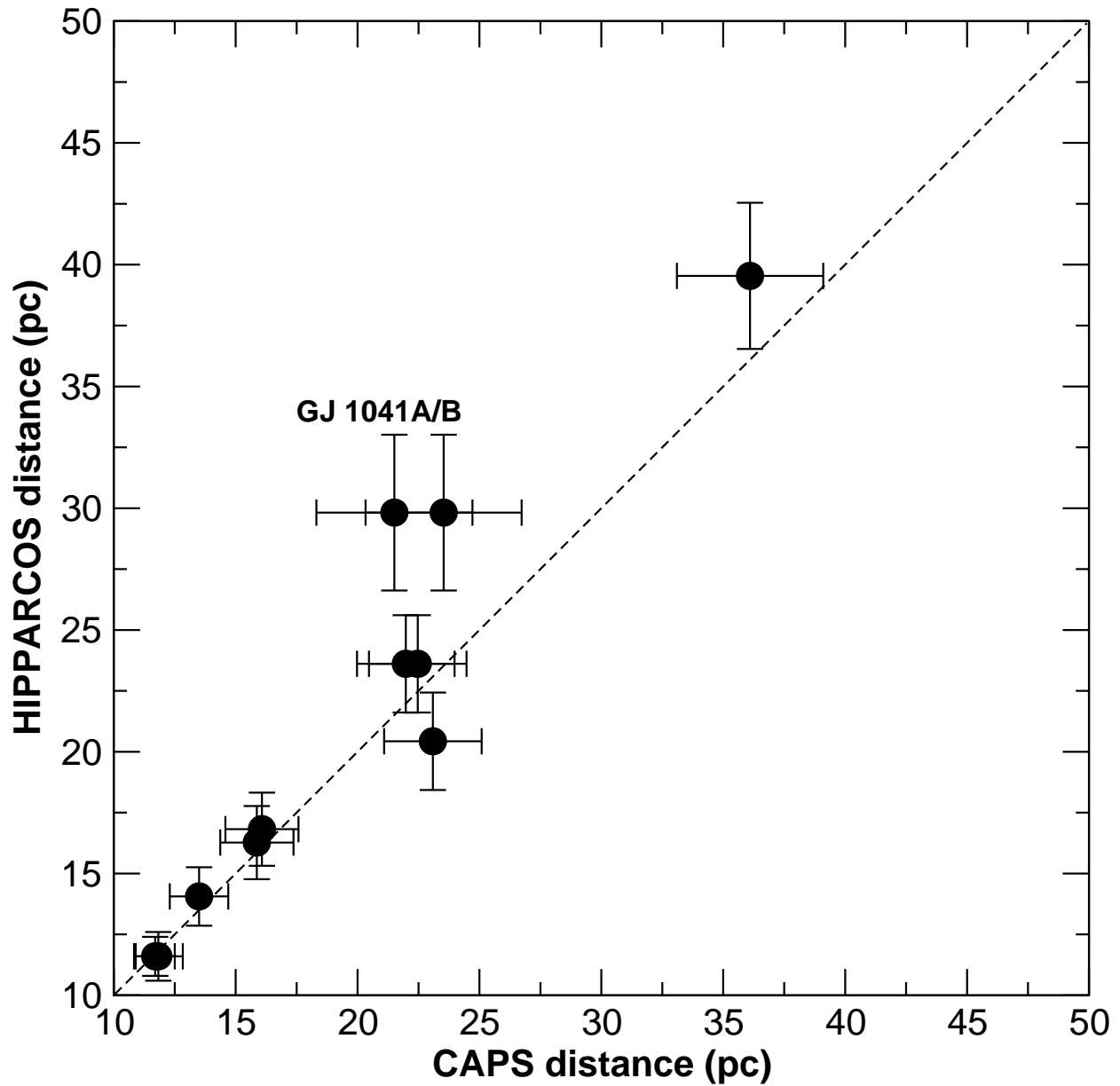}
\caption{A comparison of our trigonometric distances to stars with existing Hipparcos distances. The
agreement is consistent within the expected uncertainties except for GJ~1041AB. This is a 4$\arcsec$ binary with components of similar brightness, so both the Hipparcos and CAPSCam parallaxes might be affected by centroid systematics. Excluding GJ~1041~AB, the RMS of the differences is 1.6 pc or 10\% relative error with a $\tilde\chi^2$ of 0.888 (with 8 degrees of freedom).  The data are listed in Table~\ref{table_dist_compare}.
\label{hip_compare}}
\end{figure}

\begin{figure}
\epsscale{1.0}
\plotone{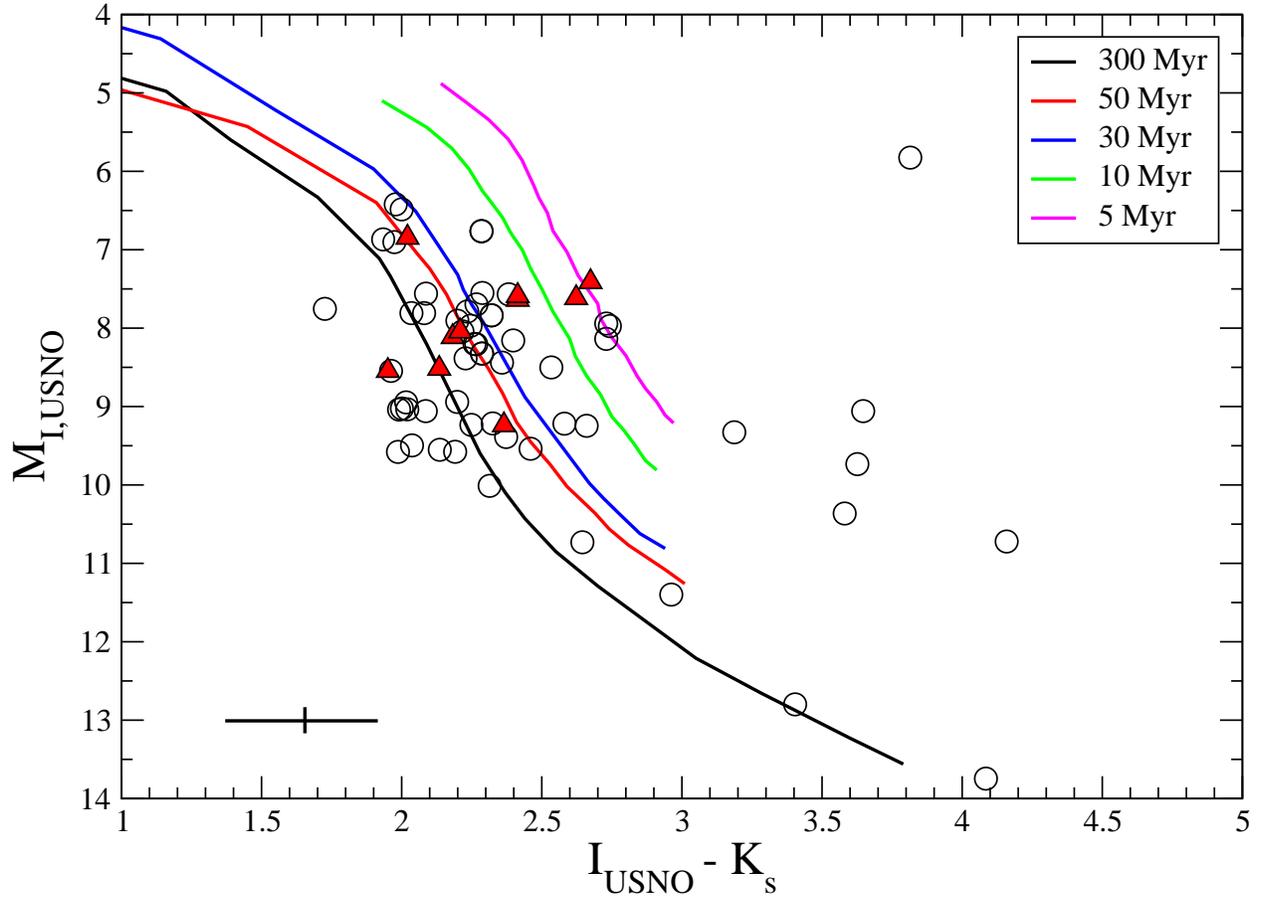}
\caption{A color-magnitude diagram of the targets with parallaxes compared to \cite{bara98} isochrones. The proposed YMG members with a `AAA' quality flag are marked as triangles. Note that stars with missing or unreliable USNO and/or 2MASS photometry are not included.
\label{cmd}}
\end{figure}

\begin{figure}
\epsscale{1.0}
\plotone{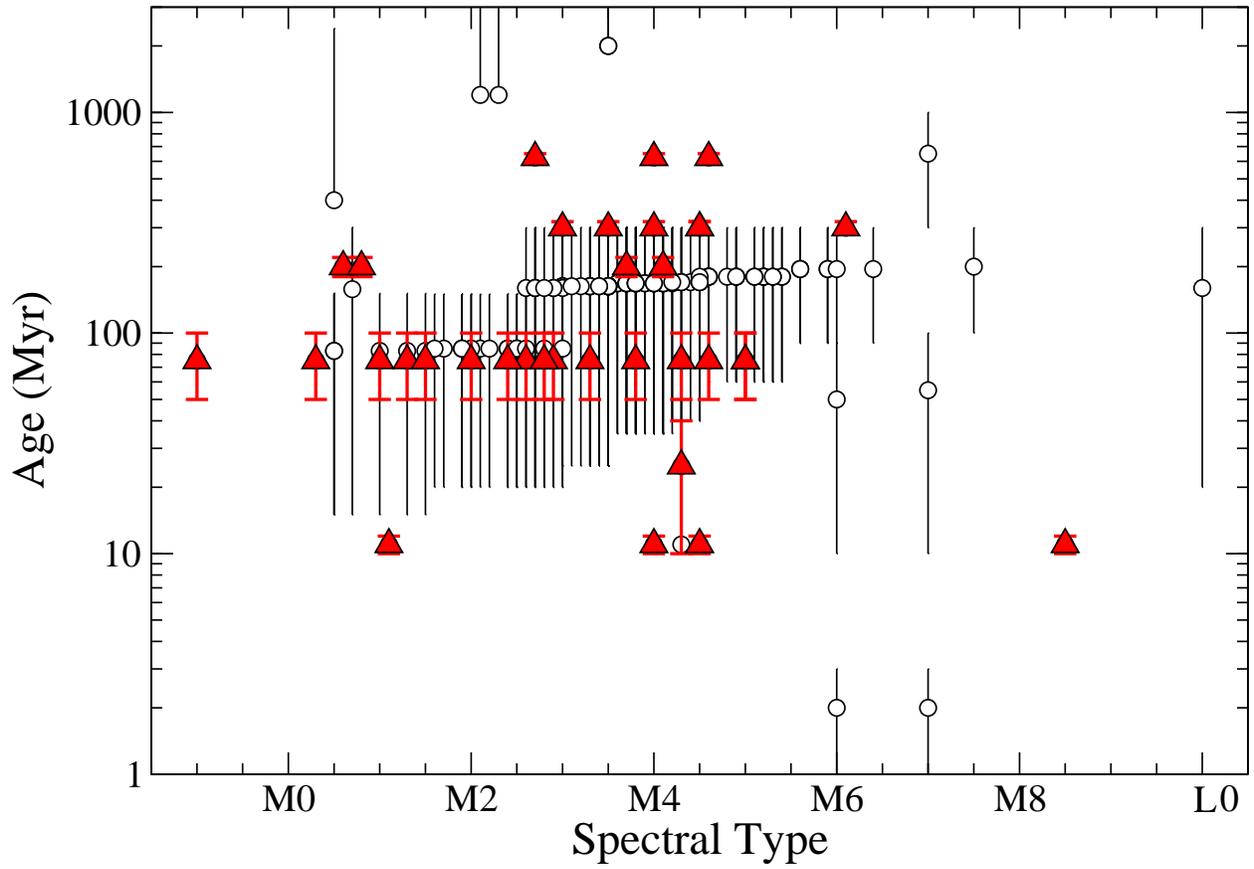}
\caption{Age distribution of our sample. The red triangles are YMG members for which the ages are more precise than those with age limits determined from the spectroscopic indicators alone.
\label{age_distribution}}
\end{figure}

Table \ref{tab:astrodata} gives an overview of the quality of the raw
astrometric measurements. In 11 cases, we can compare our trigonometric 
distances with those in the literature. As seen in Figure~\ref{hip_compare} and Table~\ref{table_dist_compare},
the agreement is fairly good,  typically well within 1-$\sigma$ level error bar,
indicating that we have been conservative in our uncertainty estimates. The RMS of the differences is 1.6 pc or 10\% relative error with a $\tilde\chi^2$ of  0.888 (8 degrees of freedom).  The color-magnitude-diagram
(CMD) of the stars with parallaxes is shown in Figure~\ref{cmd} where more than two-thirds of the
stars lie above the 300 Myr isochrone \citep{bara98}, consistent with their
overluminosity due to young ages. The age distribution of our sample is shown in Figure~\ref{age_distribution}.

\section{Stellar Kinematics}\label{kinematics}

We use our  RV measurements  in conjunction with the star's distance and proper motions to measure the 3-dimensional space velocity ($UVW$), with $U$ positive in the direction of the Galactic center, $V$ positive in the direction of Galactic rotation, and $W$ positive in the direction of the North Galactic Pole.  This provides a way to refine stellar ages by linking stars kinematically to known YMGs and associations. However, as discussed below, kinematics alone are not sufficient to assign group membership.

\subsection{$UVW$ Space Velocities}\label{uvw}

The heliocentric velocities of the stars in Galactic coordinates and their uncertainties were computed using the prescriptions given by \cite{john87}. In order to find the best kinematic match to a YMG we evaluated the
following reduced $\tilde\chi^2$ statistic with 3~degrees of freedom:

\begin{eqnarray}
\tilde\chi^2 = \frac{1}{3}\left[
\frac{\left(U_*-U_{MG}\right)^2}{(\sigma_{U*}^2+\sigma_{U,MG}^2)}+
\frac{\left(V_*-V_{MG}\right)^2}{(\sigma_{V*}^2+\sigma_{V,MG}^2)}+
\frac{\left(W_*-W_{MG}\right)^2}{(\sigma_{W*}^2+\sigma_{W,MG}^2)}
\right]
\end{eqnarray}

\noindent where $U_*$, $V_*$ and $W_*$ are the components of the heliocentric velocity of the star in Galactic coordinates and $U_{MG}$, $V_{MG}$ and $W_{MG}$ are the average components of the heliocentric velocity of
the moving group ($MG$). $\sigma_{U*}$, $\sigma_{V*}$ and $\sigma_{W*}$ are the measurement uncertainties derived
from the observations and $\sigma_{U,MG}$, $\sigma_{V,MG}$ and $\sigma_{W,MG}$ are the internal velocity scatter (i.e.~standard deviation of the velocities) of a given YMG.

\begin{figure}
\epsscale{1}
\plotone{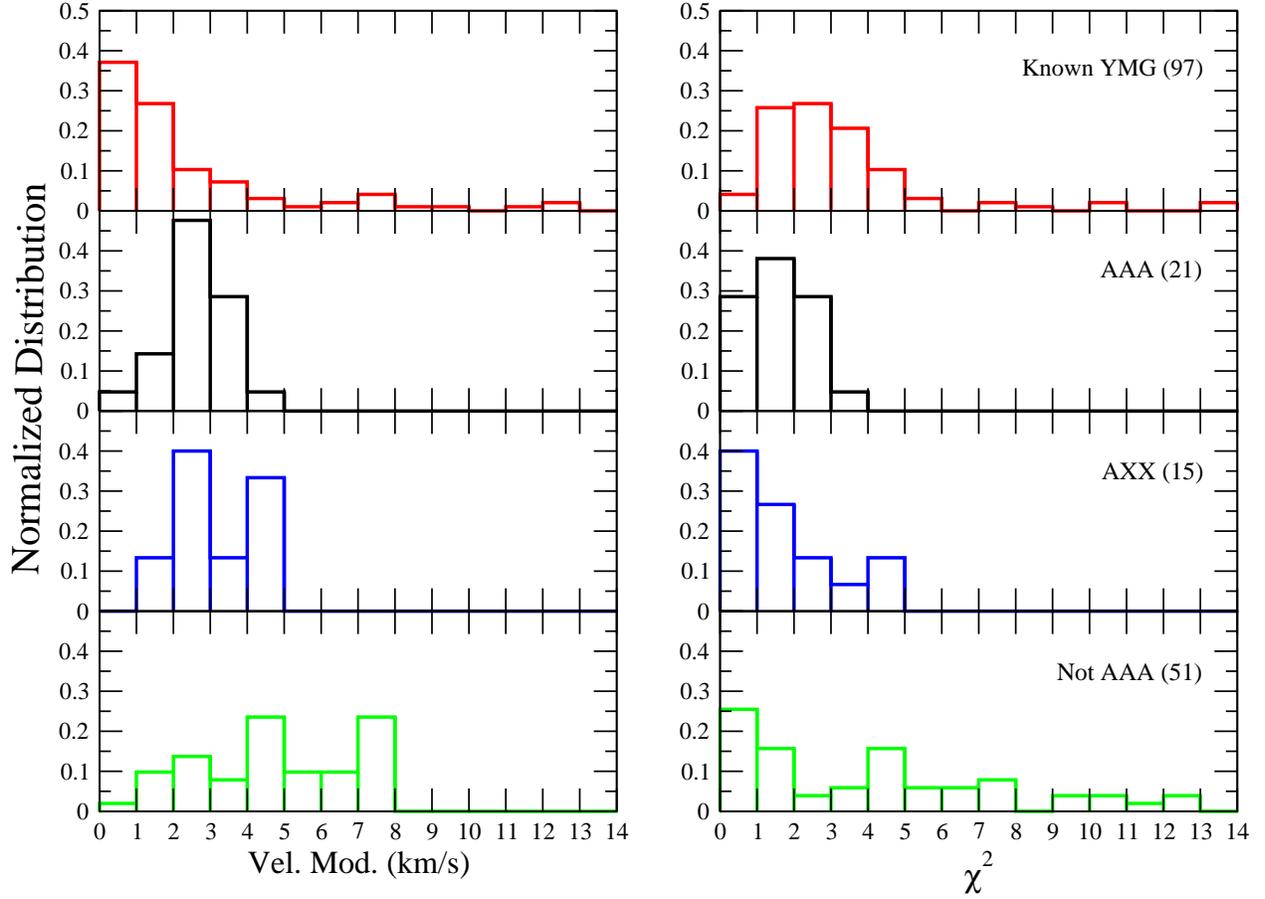}
\caption{The velocity modulus and $\tilde\chi^2$ distribution of the known YMG members (AB Dor, TucHor, and $\beta$ Pic) from \cite{zuck04} with published parallaxes. Interestingly, the ABB, ABA, and AAB members (see Section~\ref{ymg_memberships}) also have a comparable  $\tilde\chi^2$ distribution.  %The mean $\tilde\chi^2$ values are: YMG members: 3.2, AAA: 2.9, and AXX: 2.8. So, at least 
We define a good kinematic match as having velocity moduli $<$ 5 km~s$^{-1}$ and $\tilde\chi^2 <$ 6 with possible matches having $\Delta v<$ 8 km~s$^{-1}$ and $\tilde\chi^2 <$ 9.
\label{velmod_chi2_distribution}}
\end{figure}

This $\tilde\chi^2$ is a merit function that should be close to 1 if the $UVW$ of the star and
the $UVW$ of the YMG coincide at 1-$\sigma$ level in all 3 velocity components. We accept stars as kinematic matches if  $\tilde\chi^2<9$. Such a threshold ensures that we will recover 97\% of the possible members of a sample of stars.  
We note that moving groups with large internal velocity scatter will be favored if a star has a similar $UVW$ match to two associations. For matching purposes and to avoid over-weighting associations with large intrinsic scatters, we assume $\sigma_{U,MG}$ = $\sigma_{V,MG}$ = $\sigma_{W,MG} = 2$ km s$^{-1}$, corresponding to what is typically 1$\sigma$ of the velocity dispersion reported for the YMGs. Also, to avoid making wrong matches for stars with large $UVW$ uncertainties, we only accept matches when the velocity modulus ($\Delta v= \sqrt{(U_*-U_{MG})^2 + (V_*-V_{MG})^2+(W_*-W_{MG})^2}$), a measure of the difference between the $UVW$ of a star and the $UVW$ of a moving group, is less than 8 km~s$^{-1}$.  We chose an 8-km~s$^{-1}$ limit in 3-dimensional velocity space in order to find all possible matches assuming a $\approx$2~km~s$^{-1}$ velocity dispersion in each of $U, V$ and $W$ of the YMG members. Because the vast majority of the known YMG members fall within these $\tilde\chi^2$ and velocity modulus limits (Figure~\ref{velmod_chi2_distribution}), we further refine our criteria to $\Delta v<$ 5~km s$^{-1}$ and $\tilde\chi^2$ $<$ 6 to identify the most probable members. These limits were set by the $\Delta v$ and $\tilde\chi^2$ distributions of the known YMG members of which 90\% lie within these limits (Figure~\ref{velmod_chi2_distribution}).

Eighty-three of our targets have trigonometric parallaxes. For these, the $UVW$s are more precise than stars that do not have parallaxes. For the remaining targets, we calculated photometric distances using the \cite{bara98} solar-metalicity models. Our input data were effective temperatures using the \cite{legg00} and \cite{legg01} conversions from spectral types, 2MASS $K_s$ magnitudes, and the upper and lower age limits determined from the spectroscopy and X-ray emission (SLR09). The mean photometric distances are reported in Table~\ref{table_targets} with uncertainties spanning the full age range. Photometric distances are also corrected for any known binaries unresolved in the catalog photometry, either found during the spectroscopic and/or parallax programs described here or from adaptive optics imaging of these objects by Bowler et al.~(in preparation and see Section~\ref{vb}). 

\begin{figure}
\epsscale{1.0}
\plotone{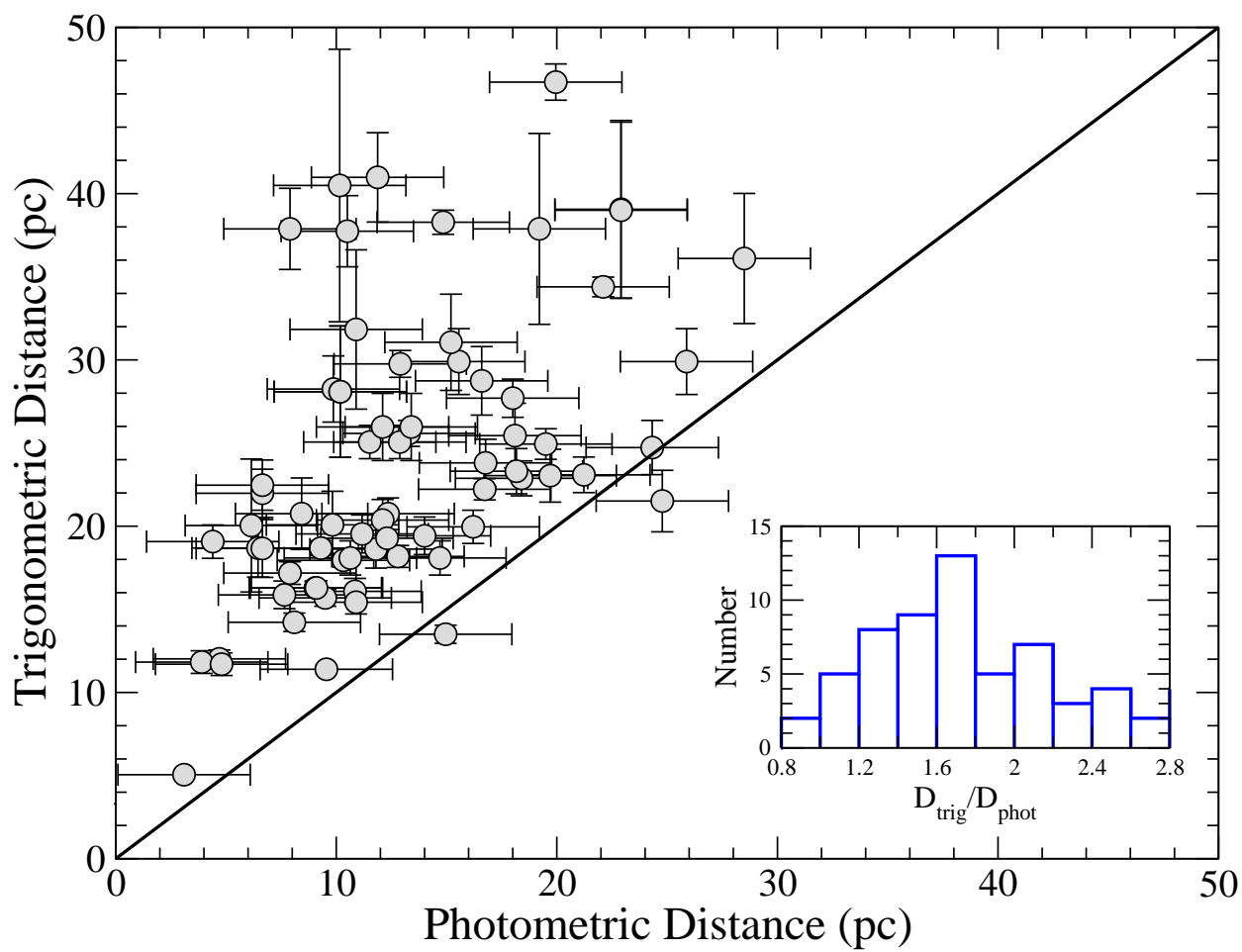}
\caption{Comparison between photometric and trigonometric distances. The photometric distances have been corrected for their youth using upper age limits and known binarity.  Note, two points are excluded for the youngest stars (2MASS J0557--1359 and 2MASS J1553--2049 ) which have distances $\gtrsim$300 pc.
\label{dist_compare}}
\end{figure}

A comparison of the trigonometric distances with the photometric distances corrected for
youth (using the upper age limits from the spectroscopic diagnostics) and
resolved binaries is shown in Figure~\ref{dist_compare}.  Excluding the two very
young stars at distances of a few hundred pc, the photometric distances
underestimate the true distances by 60--80\% on average. It is important to note that the uncertainties of the photometric distances are relatively large in part due to the range in possible ages, errors in the SpT-T$_{eff}$ conversion with T$_{eff}$ known to $\pm$100K, and unknown stellar metallicities. 

\subsection{Young Moving Group Memberships}\label{ymg_memberships}

\begin{figure}
\epsscale{1.0}
\plotone{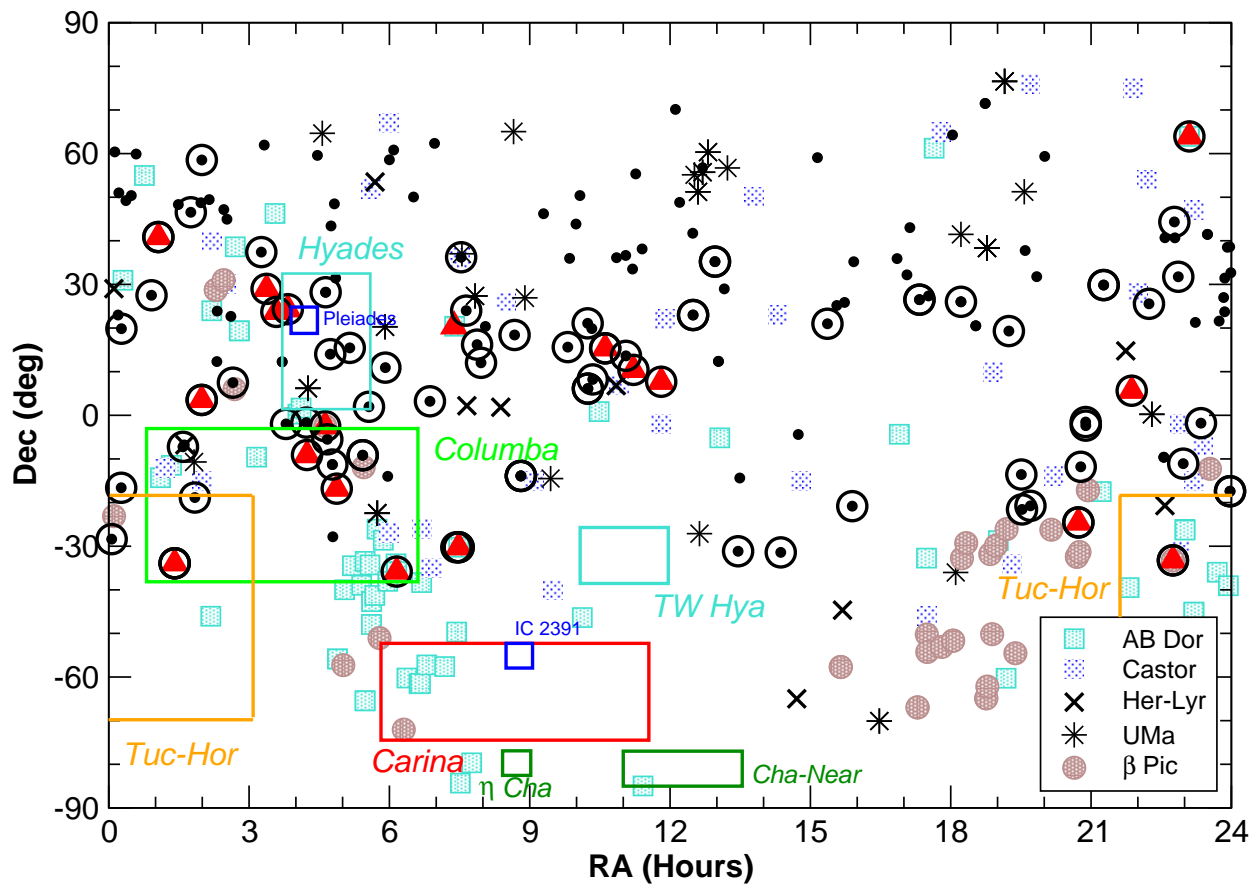}
\caption{Whole sky representation of our targets (black dots) and the YMGs.  Those with parallaxes are marked with black circles. The stars with a high-likelihood of membership to a YMG (i.e. `AAA' designation in Table~\ref{table_kinematics}) are marked with red triangles. The sources for the known members are: \cite{mont01,zuck04,lope06,lope10,torr08}.
\label{radec}}
\end{figure}

We searched for kinematic matches to 14 reported nearby YMGs listed in Table~\ref{table_ymg}. In addition to common $UVW$s, we require that the age of the YMG lie within the age range provided in SLR09 and that the 3-D position of the candidate member coincide in the sky (R.A., Dec., and distance) near the previously reported members (Figure~\ref{radec}) to propose membership to a group. Note that this procedure will favor membership in moving groups spread over the sky (e.g. $\beta$ Pic and AB Dor).

Given our cutoff criteria of $\Delta v < 8$ km~s$^{-1}$, we identify the kinematic matches in Table~\ref{table_kinematics}, each with an appended quality flag. This flag scores 3 characteristics with an `A' or `B'. These are:\\\\
(1) Kinematic matches: `A' for having a trigonometric distance with $\Delta v<$ 5~km~s$^{-1}$ and $\tilde\chi^2 <$ 6, a regime where 90\% of previously reported YMG members lie (Figure~\ref{velmod_chi2_distribution}), or `B' if the target has only a photometric distance or velocity modulus $\geq$5 km~s$^{-1}$ but $<$8 km~s$^{-1}$ and $\tilde\chi^2 \geq$6 but $<$9. For those stars with this `A' flag, we provide a kinematic membership probability in Table~\ref{table_kinematics} computed by assuming that the $\tilde\chi^2$ follows a typical $\chi^2$ distribution with 3 degrees of freedom. The probabilities of the previously-reported members are all $>$10\%. \\\\
(2) Three-dimensional spatial agreement with the YMG: `A' for lying within the 3-D bounds of the YMG, determined by drawing an R.A./Dec. box around all previously-reported members (Figure~\ref{radec}) and falling within the distance range they span (Table~\ref{table_ymg}); and `B' for not. For the loose associations, such as AB Dor, $\beta$~Pic and UMa, there are no R.A. and Dec. constraints applied but distance agreement is still necessary.\\\\
(3) Age agreement with the spectroscopically determined ages: `A' if the YMG's age falls within the range reported by SLR09 or `B' if it does not.  

We consider stars flagged with `AAA' as good YMG members.

Of the 83 M dwarfs with trigonometric distances, 36 stars have kinematics consistent with one of the known YMGs, i.e.~kinematic flag =`A'. In 13 cases, a star shares its $UVW$ with more than one YMG.  For these, we disentangled the classification using the star's sky position and/or spectroscopic age limits. In the end, we identified 21 `AAA' members, 10 of which had been previously reported, plus 17 `BAA' YMG candidates.   The $\Delta v$ and $\tilde\chi^2$ distributions of the `AAA' YMG members are very similar to that of the known YMGs with published parallaxes (Figure~\ref{velmod_chi2_distribution}).  One might expect then that those stars flagged as `ABB', `ABA', or `AAB', which match YMG kinematics within our velocity in $\tilde\chi^2$ limits might have a different $\tilde\chi^2$ distribution, skewed more towards higher values. However, at least for the 15 stars in this category, this is not the case.

\begin{figure}
\epsscale{1}
\plottwo{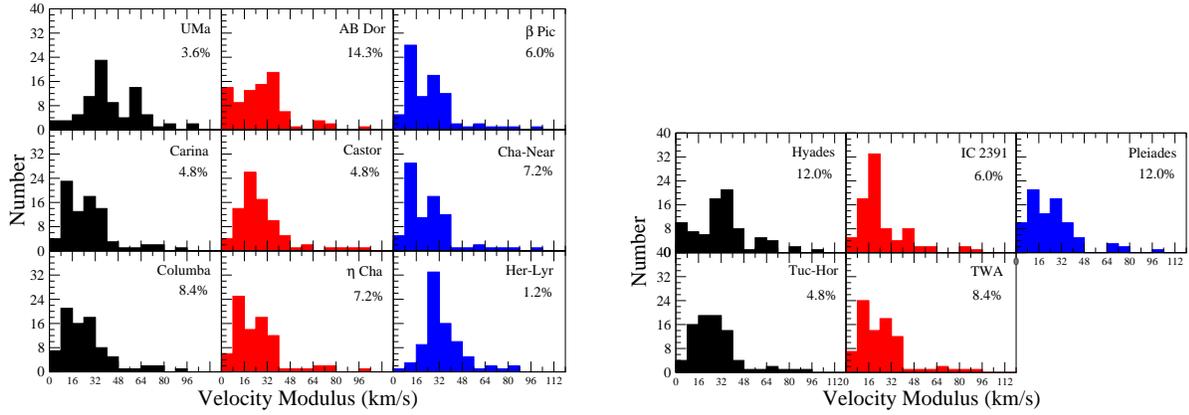}{velmod_distribution_grid2.eps}
\caption{Velocity modulus distribution of the 83 targets with parallaxes as measured for each of the 14 YMGs. The percentages listed for each YMG represents the number of kinematic matches found to that group, i.e. the fraction of targets found in the first bin ($<$8 km/s).
\label{velmod_distribution_grid1}}
\end{figure}

\begin{figure}
\epsscale{0.5}
\plotone{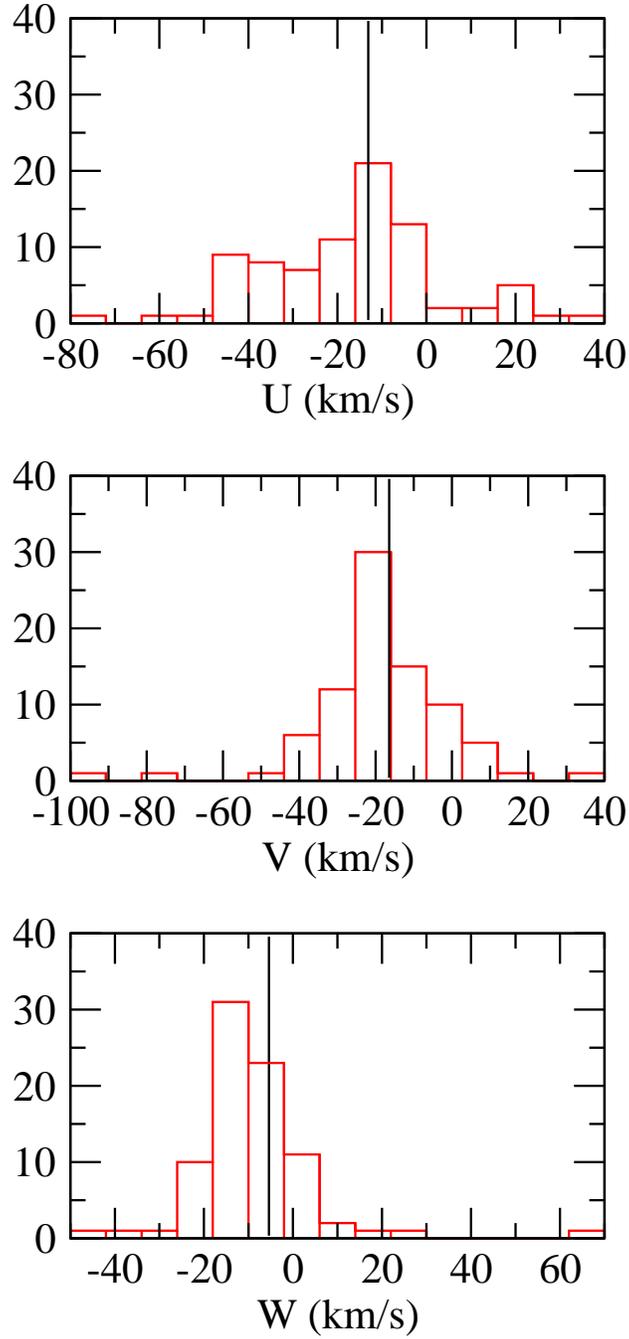}
\caption{$UVW$ distribution of the 83 targets with parallaxes. The solid line represents the mean $U$, $V$ and $W$ of the YMGs. 
\label{uvw_distribution}}
\end{figure}

\begin{figure}
\epsscale{0.5}
\plotone{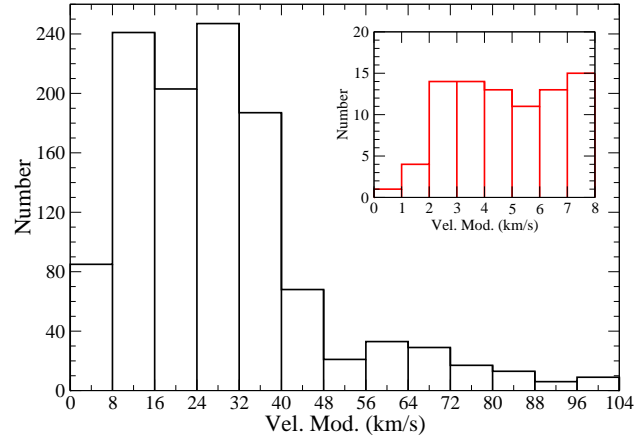}
\caption{Combined velocity modulus distribution for the stars with parallaxes as measured for all the YMG. Eighty-six kinematic matches (with $\Delta v<$ 8 km~s$^{-1}$) were found out of a possible 1162 (=83 targets $\times$ 14 YMGs) matches. A cut of 5 km~s$^{-1}$ produces 46 matches.
\label{velmod_distribution_combo}}
\end{figure}

By analyzing the kinematics of the stars with parallaxes, we find that there is on average a 4\% chance of any given star to appear as a kinematic match (flag = `A') to any one of the 14 YMGs (Figures~\ref{velmod_distribution_grid1} -- \ref{uvw_distribution}), with a total of 46 out of a possible 1162 matches (including multiple matches for a given star; Figure~\ref{velmod_distribution_combo}). Of these 46 matches, 21 are flagged as `AAA' and are very probable YMG members. This leaves us with 54\% of the matches which do not adhere to our other two criteria.   Due to this high rate of false positives of a Galactic disk star with $UVW$s matching a YMG within our confidence level, it is critical that common kinematics be only one of several criteria required for group membership.   

An example of such a pitfall is the proposed Hyades Stream defined by $\sim$30 stars with $UVW$ velocities matching those of the Hyades Cluster \citep{fama05}. Using chemical analysis, \cite{desi11} showed that the metallicity distribution of the stars in the Hyades Stream resembles that of the Galactic field stars rather than of the cluster itself, with only 4 stars matching that of the cluster. Similarly, \cite{fama07} observed the stream stars to have a mass distribution closer to the field than to the Hyades Cluster itself, which is depleted in low-mass stars. \cite{pomp11} interpreted the Hyades Stream as an inner 4:1 resonance of the Galactic spiral pattern, rather than stripped off members of the cluster.

Similarly, \cite{fama08} and \cite{bovy10} found that the bulk of proposed members of the Pleiades Moving Group, as determined from common $UVW$s, were not stripped off members of the Pleiades cluster as previously believed. These results demonstrate a valuable lesson. In order to firmly identify YMG members, a confluence of evidence is required, namely, common 3-dimensional space velocities, coherent sky position, consistent age estimates, and when practical (i.e. for FGK stars), compositional homogeneity.

To summarize our findings, we identify with reasonable confidence 
9 new members of the AB Dor YMG (50--100 Myr)  and
2 members of the Ursa Majoris group (300 Myr), bringing the count up to 43 and 30 members, respectively. We also recovered 10 previously reported YMG members.
And with lower probability, we identify 
6 possibly new members of the Castor group,
4 of the Ursa Major moving group,
2 AB Dor members,
and 1 member each of the Her-Lyr and $\beta$~Pic groups.
We also identify the first 3 members of a potentially new YMG consisting of a binary pair of M dwarfs (2MASS J04465175-1116476 A and B) and a co-moving third M dwarf (GJ 4044). (See Section~\ref{notes}.) Although they are far apart on the sky, they each have a distance of 18 pc and agree in age, both ranging from 40 -- 300 Myr.

\subsection{Where are all the `Missing' M Dwarf Members?}

\begin{figure}
\epsscale{1.0}
\plotone{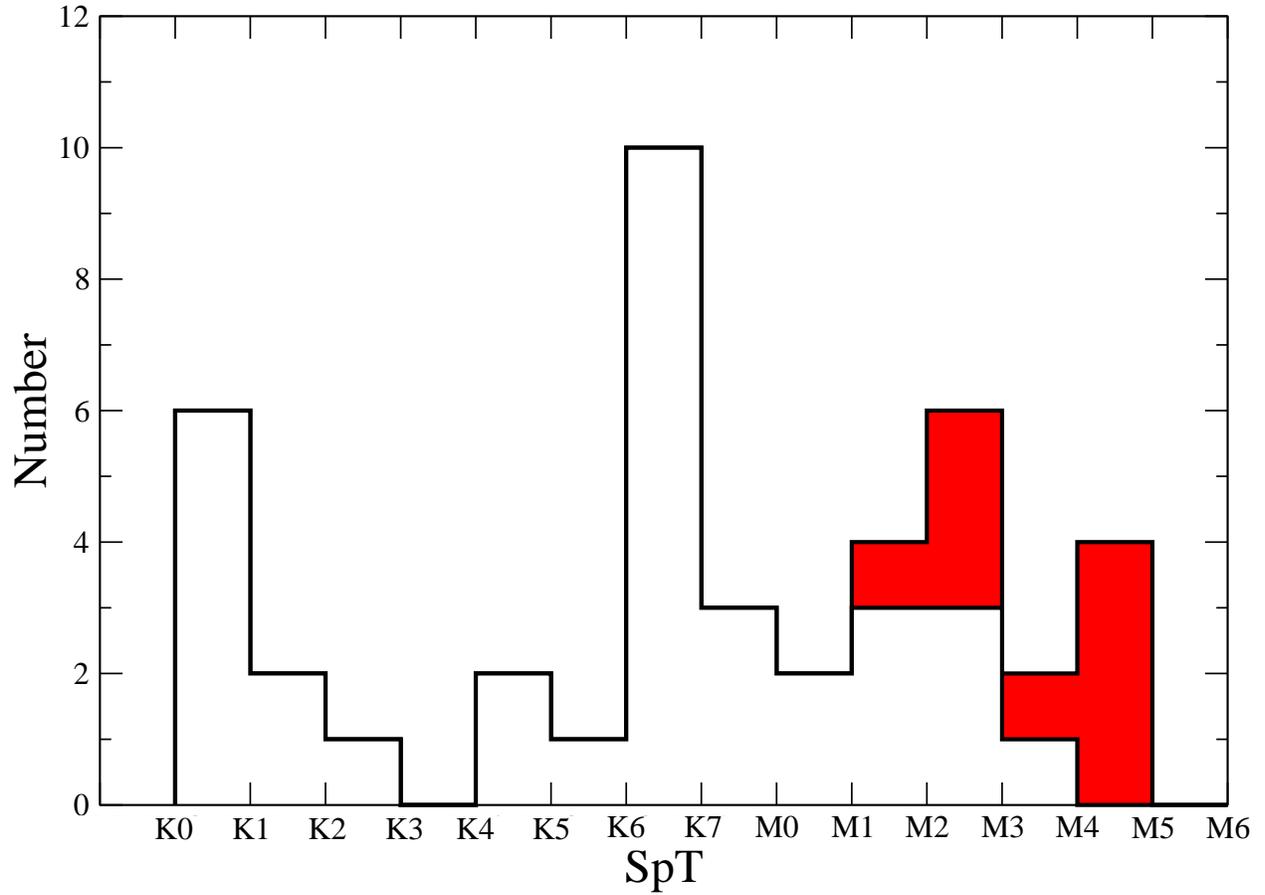}
\caption{SpT histogram of previously reported AB Dor members (black; \citealt{zuck04b,zuck11,schl10}) with our 9 proposed AB Dor members (red). The bins represent the number of stars in each spectral type range, e.g., the last bin contains those stars with SpTs $>$M4 and $\leq$M5.
\label{abdor_hist}}
\end{figure}

As noted in Section~\ref{intro}, we expect there to be many more M dwarf members of YMGs than currently known. 
Although we report a total of 25 new low-mass candidate members (with varying probabilities) of several YMGs, this is only a small fraction of the expected true population.  For example,  we expect AB Dor to have an additional $\approx$90 M dwarfs yet we found only 1/10 of that (Figure~\ref{abdor_hist}).   

\begin{figure}
\epsscale{1.0}
\plotone{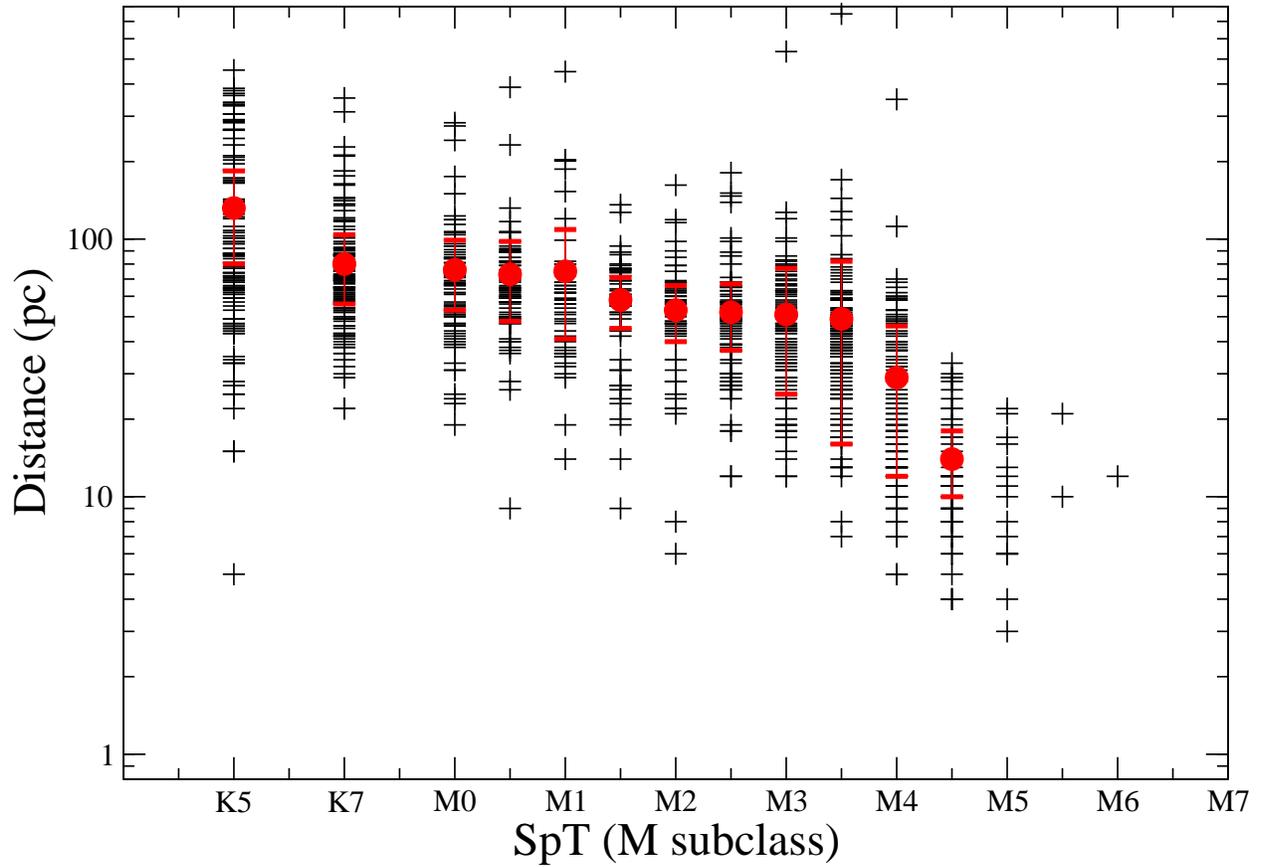}
\caption{The SpT and photometric distance distribution of the \cite{riaz06} sample of \rosat\ selected late-K and M stars. Red points are the mean values with 1 standard deviation bars.
\label{riaz06}}
\end{figure}

There are several reasons for this, the most obvious one being the selection criteria of our original SLR09 sample. We selected only targets with declinations greater than --35$^{\circ}$, i.e.~those accessible to the Mauna Kea telescopes.  This eliminated the potential of finding new members of the very southern associations such as $\eta$ Cha.  Yet we still searched for common space-motion objects to these southern YMGs in case  of any associated streams of stars. 
Another factor is \rosat's insensitivity to finding mid- to late- M dwarfs past 30~pc.
According to Figure~\ref{riaz06}, showing  the \rosat-selected sample of \cite{riaz06},  we would not be sensitive to many young stars later than M4.5. In addition, it may be that the intrinsic dispersion in $L_X/L_{Bol}$ leads us to choose only the most active members of each YMG.

It is also possible that the published FGK members of the YMGs are not all actually members thereby skewing our expected M dwarf count.   
Lastly, it may even be possible that the mass function of these YMGs strongly differs from the field population, and the M dwarf members are not actually there to be found, as appears to be the case for Taurus star-forming region where the relative number of low-mass stars appears to be deficient compared to other young clusters \citep{luhm04}.
In order to test the implications of \rosat's limitations as well as to search for more low-mass YMG members, we are completing a far more sensitive survey of young M dwarfs with photometric distances out to 100 pc using the \galex\  archive \citep{shko11} and telescopes located in both the northern and southern hemispheres.

\section{Visual Binaries}\label{vb}

Visual inspection of the images taken during the first epoch of each target revealed 10 VBs, 3 of which were previously unreported. For 8 of these,  we use the ``lucky imaging'' technique (e.g.~\citealt{law06}) to measure their separation, position angle and approximate flux ratio in the CAPSCam bandpass. (For the remaining two, 1RXS 2043 and G273-191, the lucky imaging procedure fails as the two components are of near-equal flux.) CAPSCam allows for fast readout of a small window on the detector while integrating deeply on the full field.   For standard astrometric observations, this guide window is directly added to the full field image so the perturbations of the atmosphere and tracking are the same for the target star and the background reference sources. However, since these sub-images are stored in independent files,  they can be shifted and added to produce a higher contrast image.  
This reduction process significantly enhances the contrast of the image resulting in better resolution than that typically allowed by the seeing. 

\begin{figure}
\epsscale{1.0}
\plotone{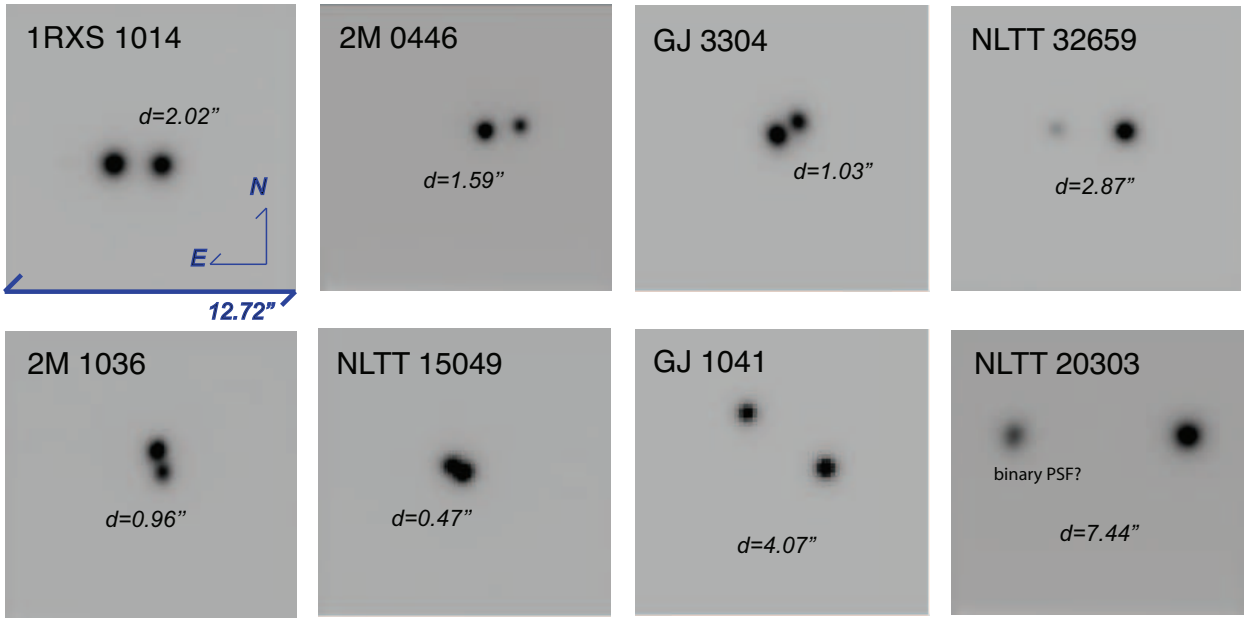}
\caption{`Lucky' images of the 8 VBs observed with CAPSCam. See text and Table~\ref{table_vb} for more details on each binary pair. 
\label{lucky}}
\end{figure}

\begin{figure}
\epsscale{0.5}
\plotone{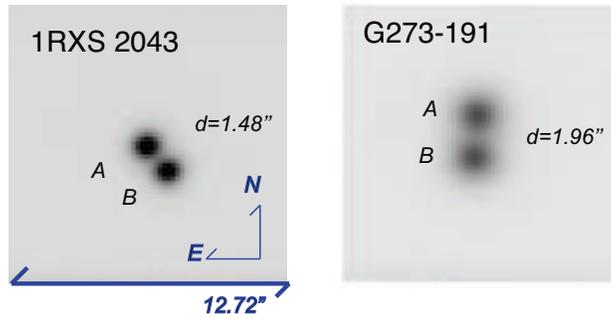}
\caption{Images of 2 VBs for which the lucky imaging technique does not work due to their near-equal fluxes. 
\label{nolucky}}
\end{figure}

When a close binary was identified, we ran this lucky imaging technique on the sub-images obtained on the first epoch of observation. Between 300 and 1500 sub-images are used to produce the final image. The results for the 10 VBs are shown in Figures \ref{lucky} and \ref{nolucky}. The angular separation, position angle and flux ratio of each pair are computed from the images and are summarized in Table~\ref{table_vb}. We also searched each field for wide common-proper motion stars to our targets and found 4 such systems, which are described in more detail below.

In addition to these relatively wide visual binaries, many stars in the SLR09 sample have been resolved into 
very tight binary systems 
through a separate high contrast adaptive optics (AO) imaging program being carried out at the Keck and Subaru telescopes
(\cite{bowl12a}, in press, and in preparation).  These binaries resolved with AO are identified with a ``(AB)'' designation in the tables and
are not resolved in our CAPSCam data.  The natural guide star AO observations were conducted 
with the Near Infrared Camera~2 (NIRC2) in its narrow field mode on Keck-II and the 
High Contrast Instrument for the Subaru Next
Generation Adaptive Optics (HiCIAO; \citealt{hoda08}) imager on Subaru.  
Table \ref{tab:aobinaries} lists the flux ratios of the resolved binaries, which were derived by fitting
analytical models made of the sum of three elliptical Gaussians for each binary component as
described in \cite{liu08b}.  The quoted values and uncertainties are the mean and standard
deviations from fits to individual images.  We found that variations in the input model (e.g., two versus
three Gaussians per binary component) resulted in systematic errors in the resulting flux
ratios of $\sim$5\% for blended binaries and $\sim$1\% for well separated systems.  We therefore
add 5\% and 1\% of the flux ratios in quadrature with the random errors we measure for binaries
with separations $<$150~mas and $>$150~mas, respectively.
More details about our observations 
will appear in an upcoming paper describing this AO imaging survey.

\section{Notes on Individual Targets}\label{notes}

\emph{2MASS J00034227--2822410}
is an M7 dwarf common proper motion companion to the G8.5V star HD 225118 lying 65.35\arcsec\ away, first noticed by \cite{cruz07}. We confirmed this companionship with mutually consistent RVs: RV$_{\mathrm{G8}}$=10.6 $\pm$ 0.3 km~s$^{-1}$ \citep{crif10} and RV$_{\mathrm{M7}}$=11.6 $\pm$ 1.1 km~s$^{-1}$. 
Using CAPSCam, we measured the distance to the G8 star to be 36.1 $\pm$ 3.9 pc, in good agreement with distance to the M7 dwarf of 38.6 $\pm$ 4.0 pc measured by Dupuy \& Liu (2012, in press).\footnote{We were unable to obtain a reliable parallax for the M dwarf due to its faintness.}
\cite{fahe10}\footnote{Note that the $UVW$s reported for the M dwarf by \cite{fahe10} are inconsistent with ours, yet the proper motions, distances and RVs generally agree.} determined that the age of the M dwarf is between 0.1 and 1 Gyr  because it does not display signs of low gravity yet is still X-ray active (\citealt{west08}). Similarly, SLR09 find no indication of low gravity in its optical  spectrum. Its strong X-ray activity implies a probable upper age limit of $\sim$300 Myr.  However, \cite{fahe10} provide an age range for the primary of 0.9 to 1.4 Gyr based on Ca II activity age \citep{mama08}.  
Given this inconsistency, the age of the system could be between 0.3 and 1 Gyr.  

\emph{G 132-51 B (E \& W)} is a common proper-motion companion to G 132-50 AB (see Table~\ref{tab:aobinaries}). Including G 132-51 A makes this a quintuple system in the AB Dor Moving Group.											

\emph{G 271-110} has recently been identified as a common-proper motion companion to EX~Cet by \cite{alon11}, a known young star reported to be part of the Local Association by \cite{mont01}. We measure the RV of G 271-110 to be 12.2 $\pm$ 0.4 km~s$^{-1}$ which is consistent with the published RV of EX Cet (11.6 $\pm$ 0.6 km~s$^{-1}$, \citealt{mont01}). We therefore adopt the Hipparcos distance of 24 pc of EX Cet for G 271-110. We calculate $UVW$ of  (-13.2, -19.1, -11.8) km~s$^{-1}$ making it a good kinematic match to the Pleiades, albeit very distant in 3-D space from the cluster.

\emph{2MASS J03350208+2342356} is an M8.5 brown dwarf. It has detected lithium absorption and signs of ongoing accretion, setting an upper age limit of 10 Myr. Its $UVW$ velocities produce $\Delta v$ of 7 km~s$^{-1}$ making it a lower-probability (8\%) `BAA' member of the $\beta$ Pic YMG . If this BD is indeed a member, then it would be equal to the lowest mass member known, 2MASS~J06085283-2753583 \citep{rice10}, which has a $\Delta v$ of $\sim$4 km~s$^{-1}$.

\begin{figure}
\epsscale{0.5}
\plotone{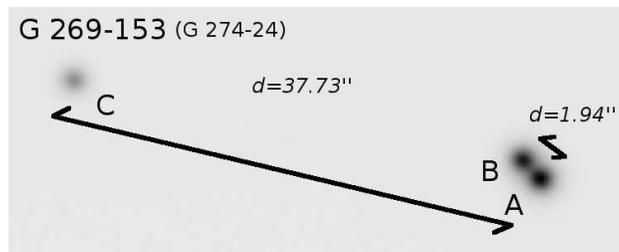}
\caption{Image of G 269-153 ABC observed on UT20090909. 
\label{triple}}
\end{figure}

\emph{G 269-153 NE, SW, E = GJ 2022 ABC:} 
RVs for these three AB Dor members agree, and we adopt a trigonometric distance of 12.6 $\pm$ 2.3 pc for the system based on our parallax for the brightest component. At the times of our observations, between UT20090909 (Figure~\ref{triple}) and UT20101114, the NE (SpT=M4.3) and SW (SpT=M4.6) components were $<1\arcsec$ apart, closer than the 2.0$\arcsec$ reported by \cite{daem07} in both 1999 and 2005. This difference is likely due to the orbital motion of the binary.  
Since the C component, lying 38\arcsec\ eastward, was not included in the original SLR09 sample, we observed it with the MIKE high-resolution optical spectrograph at the Magellan (Clay) Telescope. We measured its RV to be 18.24 $\pm$ 0.45 km~s$^{-1}$ and its SpT as M5.

\emph{GJ 3305 AB} (SpT = M1.1)
is resolved as a tight binary by high-resolution imaging by \cite{delo12} and Bowler et al.~(in preparation), but not by us using CAPSCam.  It is a clear common proper motion companion to the $\beta$ Pic YMG member and F0V star 51~Eri (HD 29391) separated by 66.7\arcsec\ \citep{feig06}.  The RV of 51~Eri is 21.0 $\pm$ 1.8 km~s$^{-1}$ \citep{khar07}, which agrees with that of GJ~3305~AB (21.7 $\pm$ 0.3 km~s$^{-1}$).  Thus we use 51 Eri's Hipparcos distance of 29.8 $\pm$ 0.8 pc to calculate the $UVW$ for GJ~3305~AB.  At this distance, the separation between 51 Eri and GJ 3305 AB is 2200 AU.

\emph{NLTT 13728} is a newly confirmed young M6 brown dwarf within 10 pc. We measured its distance to be 9.5 $\pm$ 0.3 pc with an upper age limit of 90 Myr based on signatures of low-gravity and strong H$\alpha$ emission (SLR09).

\emph{2MASS J04465175--1116476 AB}:
is a binary pair with a separation of 1.59\arcsec\ and a position angle of 279$^{\circ}$. The primary (east component; SpT=M4.9) is 1.93 times brighter than the secondary (SpT=M6) in the CAPSCam bandpass (roughly I band). 
The binary and its co-moving companion GJ 4044 have distances of 18.6 $\pm$ 1.7 pc and 17.2 $\pm$ 0.7 pc, respectively, and $UVW$s of (-5.3	$\pm$	0.7,	-0.9	$\pm$	1.2,	-20.0	$\pm$	1.2) and (--4.7	$\pm$	0.5,	0.3	$\pm$	0.5,	-19.5	$\pm$	0.8) km~s$^{-1}$. 
Although they are far apart on the sky (separated by 212$^{\circ}$), they do agree in age, both ranging from 40 -- 300 Myr.

\emph{2MASS J05575096--1359503} is a young M7 object with a photometric distance of 60 pc using an age of 10 Myr and the \cite{bara98} models.  However, the parallax is so small that the distance to this object is formally measured to be 526 $\pm$ 277 pc. This larger distance implies that the object is a brown dwarf younger than 2 Myr.

\emph{NLTT 15049} is another tight VB with the primary (SpT=M3.8) lying SW of the secondary (SpT=M5).  The separation is 0.47\arcsec\ with a position angle of 56$^{\circ}$ and a CAPSCam flux ratio of $\approx$3.

\emph{CD-35 2722} is an M1 star reported to have a wide substellar companion by \cite{wahh11}. Our kinematic and age analysis identifies this system as a highly-probable (98\%) AB~Dor member, agreeing well with the results of \cite{wahh11}.

\emph{1RXS J101432.0+060649 AB} is a new VB with the primary (M4.1) lying east of the secondary (M4.5) with a separation of 2.02\arcsec, a position angle of 271$^{\circ}$, and a CAPSCam flux ratio of 1.2.

\emph{2MASS J10364483+1521394 AB} is a triple system where the northern component (A; SpT=M4) is 1.7 times brighter in the CAPSCam bandpass than the southern component (B; SpT = M4.5+M4.5) with a separation of 0.96\arcsec\ and a position angle of 160$^{\circ}$. \cite{daem07} reported a separation of the components of 2MASS J10364483+1521394 B to be 0.189\arcsec, undetectable by us as this separation corresponds to about one CAPSCam pixel. 

\emph{NLTT 56194} has a SpT of M7.5 and a photometric distance of 14 $\pm$ 3 pc with a spectroscopic age of 100--300 Myr.  It is a kinematic match, with a quality flag of `B', to the Castor group (200 Myr). This make NLTT 56194 another young BD in our sample.

\section{Summary}\label{summary}

We present a kinematic analysis of the 165 young ($\lesssim$300 Myr), nearby and X-ray-bright M dwarfs compiled by SLR09. We have further characterized this ``25-pc sample'' with radial velocities and  distances, showing that nearly half are outside of 25 pc 
as expected if they are indeed young. 
We acquired distances for half the sample from trigonometric parallaxes while we estimated photometric distances for the remainder of the sample using models and correcting for visual binarity and youth determined from spectroscopic age indicators. 
Our astrometric survey showed that for this sample of young stars, photometric distances underestimate the true distances by an average of roughly 70\%, and range from 5 to 500 pc. This implies that the $UVW$s of young stars based on photometric distances that assume the stars lie on the main-sequence are most likely unreliable.

By combining our RVs and distances with proper motions, we calculated the targets' $UVW$ velocities in search of new YMG members. We identified 21 likely members of YMGs, which have accurate $UVW$s determined from trigonometric parallaxes, positional coincidence with known members, and age agreement using the independent spectroscopic and X-ray age diagnostics presented in SLR09. Of these, 10 were previously known AB Dor, Hyades, or $\beta$~Pic members. Newly proposed members include 9 AB Dor members and 2 UMa members. Additional possible candidates include
6  Castor members,
4 Ursa Majoris members,
2 AB Dor members,
and 1 member each of the Her-Lyr and $\beta$~Pic groups.

In addition, we identified three stars (two of which form a binary pair) with identical kinematics and age ranges, potentially forming their own young association. We also found kinematic matches to YMGs for 40 stars, but either their positions on the sky or ages do not agree with previously known members. For 12 stars that are kinematic and age matches but are positionally offset to the other group members, chemical analysis would be one way to confirm that the stars did indeed originate from a common star-forming event. However, metallicity indicators of M dwarfs may not yet be accurate or precise enough to discriminate group members from the field population (e.g.~\citealt{neve11} and references therein). Lastly, our sample also contains 31 young M dwarfs, including 4 brown dwarfs, with ages $\lesssim$150 Myr, which are not associated with any known YMG. 

Even relaxing the requirements of spectroscopic and positional age agreement, we did not recover the expected numbers of M dwarf members to the YMGs. After accounting for our observing limitations, the likeliest explanation is that the \rosat\ All-Sky Survey is not sensitive enough to detect YMG members lying beyond 25--30 pc. In order to understand this further, as well as identify even more members, we are completing a larger survey of young M dwarfs with photometric distances out to 100 pc using the more sensitive \galex\ archive.

\acknowledgements

We appreciate the spectrum of G~269-153 (E) taken by Julio Chaname and CAPSCam data observed by Stella Kafka. We would like to thank the CFHT,
Keck and LCO staff for their care in setting up the instruments and support in the control rooms, and to J.-F. Donati for making {\it Libre-ESpRIT}
available to CFHT users. We would also like to thank the referee, S\'ebastien L\'epine, for a careful review of the manuscript. Funding from the Carnegie Institution of Washington for E.S. and G.A.-E.~is gratefully acknowledged. Funding for M.C.L.
and B.P.B. has in part been provided by NSF grant AST09-09222 and NASA grant NNX11AC31G. We also acknowledge partial
support for the LCO observing by the NASA Astrobiology Institute under cooperative
agreement NNA09DA81A. This publication makes use of data products from the Two
Micron All Sky Survey, which is a joint project of the University of Massachusetts and the Infrared Processing and Analysis Center/California
Institute of Technology, funded by the National Aeronautics and Space Administration and the National Science Foundation.

\clearpage
\bibliography{refs_master}{}
\bibliographystyle{apj}

%%%%%FIGURES%%%%%%

\clearpage 

%TABLES for ROSAT II PAPER

\begin{deluxetable}{llcccccccccccc}																																						
\tabletypesize{\scriptsize}																																						
\rotate																																						
\tablecaption{Observed M Dwarfs\label{table_targets}}																																						
\tablewidth{0pt}																																						
\tablehead{																																																																							
																																	
\colhead{Name\tablenotemark{a}}	&	\colhead{RA \& DEC (2MASS)}						&	\colhead{SpT\tablenotemark{b}}	&	\colhead{$I_{USNO}$}	&	\colhead{$I-K_s$}	&	\colhead{$\mu_{RA}$}			&	\colhead{$\mu_{DEC}$}			&	\colhead{$\mu$\tablenotemark{c}}	&	\colhead{RV}			&	\colhead{D$_{phot}$\tablenotemark{d}}			&	\colhead{D$_{trig}$}			\\
\colhead{}	&	\colhead{J2000}						&	\colhead{M--}	&	\colhead{mag}	&	\colhead{mag}	&	\colhead{mas~yr$^{-1}$}			&	\colhead{mas~yr$^{-1}$}			&	\colhead{Ref.}	&	\colhead{km~s$^{-1}$}			&	\colhead{pc}			&	\colhead{pc}			
}																																				
\startdata																																				
																																				
2MASS J00034227-2822410	&	00	03	42.28	-28	22	41.0	&	7.0	&	15.7	&	3.7	&	266	$\pm$	10	&	-134	$\pm$	10	&	1	&	11.6	$\pm$	1.1	&	31	$\pm$	3	&	36.1	$\pm$	3.9	\\
G 217-32 (AB)                	&	00	07	42.64	+60	22	54.3	&	3.8	&	10.3	&	2.3	&	342	$\pm$	5	&	-29	$\pm$	5	&	2	&	2.7	$\pm$	0.6	&	14	$\pm$	3	&	--			\\
LP 348-40	&	00	11	53.03	+22	59	04.8	&	3.5	&	10.2	&	2.2	&	146	$\pm$	7.5	&	-229	$\pm$	5	&	2	&	-45.2	$\pm$	0.4	&	10	$\pm$	3	&	--			\\
NLTT 614                	&	00	12	57.17	+50	59	17.3	&	6.4	&	13.7	&	3.1	&	252	$\pm$	5	&	32	$\pm$	5	&	2	&	3.0	$\pm$	0.3	&	16	$\pm$	3	&	--			\\
1RXS J001557.5-163659 (AB)	&	00	15	58.08	-16	36	57.9	&	4.1	&	10.5	&	2.6	&	-128	$\pm$	10	&	16	$\pm$	10	&	3	&	21.0	$\pm$	0.8	&	13	$\pm$	3	&	18.0	$\pm$	0.4	\\
GJ 1006 A 	&	00	16	14.56	+19	51	38.6	&	3.6	&	--	&	--	&	714	$\pm$	11	&	-757	$\pm$	8	&	4	&	-28.3	$\pm$	0.3	&	8	$\pm$	3	&	20.0	$\pm$	4.0	\\
GJ 3030                 	&	00	21	57.81	+49	12	38.0	&	2.4	&	10.1	&	1.9	&	204	$\pm$	7.5	&	-31	$\pm$	6.7	&	2	&	-10.8	$\pm$	0.2	&	25	$\pm$	9	&	--			\\
GJ 3036 (AB)\tablenotemark{e}                	&	00	28	53.92	+50	22	33.0	&	3.7	&	10.3	&	2.3	&	423	$\pm$	7.5	&	124	$\pm$	6.7	&	2	&	6.3	$\pm$	0.6	&	18	$\pm$	5	&	--			\\
\enddata
\end{deluxetable}

\begin{deluxetable}{lccccccc}
	\tabletypesize{\scriptsize}																																						
\rotate																																						
\tablecaption{Summary of Astrometric Observations 
\label{tab:astrodata}}
\tablehead{
\colhead{Name}&
\colhead{First epoch (UT)}&
\colhead{Last epoch (UT)}&
\colhead{N$_{\rm epochs}$}&
\colhead{$\pi_{\rm rel}$}&
\colhead{$\pi_{\rm abs}$}&
\colhead{$\sigma_{\rm abs}$}&
\colhead{Comments}\\
\colhead{ }&
\colhead{YYYYMMDD}&
\colhead{YYYYMMDD}&
\colhead{ }&
\colhead{mas }&
\colhead{mas }&
\colhead{mas }&
\colhead{ }
}
\tablewidth{0pt}
\startdata
2MASS J00034227-2822410		&	20090909	&	20100114	&	5	&	27.2	&	27.7	&	3.0	& PM Pair to HIP 296	\\
1RXS J001557.5-163659 (AB)		&	20090908	&	20100818	&	4	&	55.2	&	55.7	&	1.3	&	-		\\
G 69-32                   	&	20090909	&	20101114	&	4	&	38.6	&	39.1	&	1.2	&	-		\\
G 269-153 A (NE)             	&	        	&	        	&		&		&	39.9\tablenotemark{a}	&	1.6	& Plx from C		\\
G 269-153 B (NE)             	&			&			&		&		&	39.9\tablenotemark{a}	&	1.6	& Plx from C		\\
G 269-153 C (E)			&	20090909	&	20101114	&	5	&	39.4	&	39.9	&	1.6	&	-		\\
2MASS J01351393-0712517		&	20090909	&	20101116	&	4	&	25.9	&	26.4	&	1.7	&	-		\\
1RXS J015027.1-185134   	&	20090909	&	20101116	&	4	&	30.0	&	30.5	&	2.5	&	-		\\
\enddata
\tablenotetext{a}{Binary PSF produced a larger than normal error in parallax.  We set a conservative uncertainty on these of 5 mas. See text for more details.}

%\tablenotetext{a}{I NEED TO PUT THESE IN SOME ORDER, RA?}

\end{deluxetable}

\begin{deluxetable}{lcccccrrclccccc}									
\tabletypesize{\scriptsize}									
%\rotate									
\tablecaption{Trigonometric Distances: CAPSCam vs. Hipparcos \label{table_dist_compare}}									
\tablewidth{0pt}									
\tablehead{									
\colhead{Name}	&	\colhead{CAPSCam}			&	\colhead{Hipparcos}			\\
\colhead{}	&	\colhead{Distance}			&	\colhead{Distance}			\\
\colhead{}	&	\colhead{pc}			&	\colhead{pc}			
}									
\startdata									
									
2MASS J00034227-2822410	&	36.10	$\pm$	3.91	&	39.54	$\pm$	1.74	\\
GJ 1041A (SW)	&	21.51	$\pm$	1.85	&	29.82	$\pm$	4.71	\\
GJ 1041Bab (NE)	&	23.53	$\pm$	0.83	&	29.82	$\pm$	4.71	\\
G 80-21                  	&	15.87	$\pm$	0.83	&	16.27	$\pm$	0.80	\\
GJ 207.1	&	16.08	$\pm$	0.78	&	16.82	$\pm$	1.10	\\
GJ 316.1	&	13.50	$\pm$	0.55	&	14.06	$\pm$	0.20	\\
GJ 2079 (AB)	&	23.09	$\pm$	1.07	&	20.43	$\pm$	0.71	\\
GJ 699B (W)	&	11.83	$\pm$	0.68	&	11.60	$\pm$	0.81	\\
GJ 669A (E )	&	11.70	$\pm$	0.68	&	11.60	$\pm$	0.81	\\
LP-984-92 	&	21.98	$\pm$	1.45	&	23.61	$\pm$	1.88	\\
LP 984-91 A (NW)	&	22.47	$\pm$	1.51	&	23.61	$\pm$	1.88	\\

\enddata									
									
\end{deluxetable}

\begin{deluxetable}{llllllllcrcrclccccc}
\tabletypesize{\scriptsize}
\rotate
\tablecaption{Young Moving Groups Searched for New Members\label{table_ymg}}
\tablewidth{0pt}
\tablehead{
\colhead{Group}	&	\colhead{$U$}	&	\colhead{$V$}	&	\colhead{$W$}	&	\colhead{Age}	& \colhead{Dist.} &	\colhead{Ref.}	\\
\colhead{}	&	\colhead{km~s$^{-1}$}	&	\colhead{km~s$^{-1}$}	&	\colhead{km~s$^{-1}$}	&	\colhead{Myr}	& \colhead{pc} &	\colhead{}
}	
\startdata											
AB Dor Moving Group	&	-8	&	-27	&	-14	&	50--100	&	15--55	&	\cite{zuck04b,lope06}	\\
$\beta$ Pictoris Moving Group	&	-11	&	-15	&	-9.5	&	10--12	&	10--50	&	\cite{torr06,torr08}	\\
Carina Association	&	-10.2	&	-23	&	-4.4	&	30	&	50--120	&	\cite{torr08}	\\
Castor Moving Group	&	-10.7	&	-8	&	-9.7	&	200	&	$\lesssim$45	&	\cite{mont01}	\\
Chamaeleon-Near Moving Group	&	-11	&	-16	&	-8	&	10	&	60--110	&	\cite{zuck04}	\\
$\eta$ Chamaeleon Cluster	&	-12	&	-19	&	-10	&	8	&	$\sim$97	&	\cite{zuck04}	\\
Columba Association	&	-13.2	&	-21.8	&	-5.9	&	30	&	50--110	&	\cite{torr08}	\\
Hercules-Lyra Association	&	-15.4	&	-23.4	&	12	&	200	&	$\lesssim$25	&	\cite{lope06}	\\
Hyades Open Cluster	&	-39.7	&	-17.7	&	-2.4	&	650	&	$\sim$47	&	\cite{ster95}	\\
IC2391 Super Cluster	&	-20.6	&	-15.7	&	-9.1	&	35--55	&	$\sim$155	&	\cite{barr99}	\\
Pleiades Open Cluster	&	-11.6	&	-21	&	-11.4	&	120	&	120--140	&	\cite{stau98}	\\
Tucana/Horologium	&	-11	&	-21	&	0	&	30	&	25--80	&	\cite{torr08}	\\
TW Hydrae Association	&	-12	&	-19	&	-8	&	8--10	&	20--120	&	\cite{ment08,torr08}	\\
Ursa Majoris Moving Group 	&	14.9	&	1	&	-10.7	&	300	&	$\sim$25	&	\cite{zuck04,king05}	\\
\enddata

\end{deluxetable}

\begin{deluxetable}{lcccccccccccccc}																									
\tabletypesize{\scriptsize}																									
\rotate																									
\tablecaption{Kinematics and Young Moving Group Memberships\label{table_kinematics}	}																								
\tablewidth{0pt}																									
\tablehead{																																																
																								
\colhead{Name}	&	\colhead{SpT}	&	\colhead{Dist\tablenotemark{a}}	&	\colhead{$U$}			&	\colhead{$V$}			&	\colhead{$W$}			&	\colhead{$\Delta v$\tablenotemark{b}}	&	\colhead{$\tilde\chi^2$}	&	\colhead{Prob. of }	&	\colhead{Kin.\tablenotemark{c}}	&	\colhead{YMG}	&	\colhead{Ages\tablenotemark{d} from}	&	\colhead{Quality\tablenotemark{e}}	\\
\colhead{}	&	\colhead{}	&	\colhead{}	&	\colhead{}			&	\colhead{}			&	\colhead{}			&	\colhead{}	&	\colhead{}	&	\colhead{Kinematic}	&	\colhead{YMG}	&	\colhead{Age}	&	\colhead{SLR09}	&	\colhead{of}	\\
\colhead{}	&	\colhead{M--}	&	\colhead{}	&	\colhead{km~s$^{-1}$}			&	\colhead{km~s$^{-1}$}			&	\colhead{km~s$^{-1}$}			&	\colhead{km~s$^{-1}$}	&	\colhead{}	&	\colhead{Match}	&	\colhead{Match}	&	\colhead{Myr}	&	\colhead{Myr}	&	\colhead{Match}	
}																															
\startdata																												
																															
2MASS J00034227-2822410	&	7.0	&	$\pi$	&	-27.2	$\pm$	4.8	&	-40.3	$\pm$	3.6	&	-19.2	$\pm$	1.4	&	--	&	--	&	--	&	--	&	--	&	300-1000	&	--	\\
G 217-32 (AB)                	&	3.8	&	phot	&	-20.8	$\pm$	4.3	&	-8.2	$\pm$	2.3	&	-5.6	$\pm$	0.9	&	--	&	--	&	--	&	--	&	--	&	35-300	&	--	\\
LP 348-40	&	3.5	&	phot	&	11.6	$\pm$	2.3	&	-41.3	$\pm$	1.9	&	19.1	$\pm$	2.5	&	--	&	--	&	--	&	--	&	--	&	$\gtrsim$2000	&	--	\\
NLTT 614                	&	6.4	&	phot	&	-18.8	$\pm$	3.1	&	-6.5	$\pm$	1.7	&	-0.9	$\pm$	0.8	&	--	&	--	&	--	&	--	&	--	&	90-300	&	--	\\
1RXS J001557.5-163659 (AB)	&	4.1	&	$\pi$	&	9.1	$\pm$	0.9	&	11.2	$\pm$	0.9	&	-18.8	$\pm$	0.8	&	--	&	--	&	--	&	--	&	--	&	35-300	&	--	\\
GJ 1006 A 	&	3.6	&	$\pi$	&	-21.1	$\pm$	13.2	&	-91.4	$\pm$	10.2	&	-42.2	$\pm$	10.7	&	--	&	--	&	--	&	--	&	--	&	35-300	&	--	\\
GJ 3030                 	&	2.4	&	phot	&	-15.4	$\pm$	7.9	&	-21.9	$\pm$	4.6	&	-3.7	$\pm$	1.8	&	5.3	&	0.5	&	--	&	 Car	&	30	&	20-150	&	BBA	\\
GJ 3036 (AB)              	&	3.7	&	phot	&	-34.9	$\pm$	9.0	&	-11.1	$\pm$	5.4	&	6.0	$\pm$	3.1	&	--	&	--	&	--	&	--	&	--	&	35-300	&	--	\\
\enddata
\end{deluxetable}

\begin{deluxetable}{lcccccclccccc}									
\tabletypesize{\scriptsize}									
%\rotate									
\tablecaption{CAPSCam Visual Binaries\label{table_vb}}									
\tablewidth{0pt}									
\tablehead{									
\colhead{VB Name}	&	\colhead{Flux Ratio}	&	\colhead{Sep.}	&	\colhead{P.A.}	&	\colhead{Note}	\\
\colhead{}	&	\colhead{in $\approx$I}	&	\colhead{\arcsec}	&	\colhead{deg.}	&	\colhead{}	\\
\colhead{}	&	\colhead{$\pm$ 5\%}	&	\colhead{$\pm$ 0.04}	&	\colhead{$\pm$ 1$^{\circ}$}	&	\colhead{}	
}									
\startdata									
									
G 269-153 AB\tablenotemark{a}	&	1.1	&	1.94	&	45	&	Observed UT20090909	\\
~~~~~~~\arcsec	&	1.1	&	2.05	&	48.9	&	Observed UT20101114	\\
G 269-153 AC	&	2.58	&	37.7	&	78.4	&	Observed UT20101114	\\
GJ 1041 ABab	&	1.29	&	4.07	&	55	&	B component is a SB \citep{shko10}.	\\
GJ 3304 AB	&	1.5	&	1.03	&	305	&	\cite{daem07} reports sep=0.78\arcsec and PA = 300$^{\circ}$.	\\
2M 0446 AB	&	1.93	&	1.59	&	279	&		\\
NLTT 15049 AB	&	3	&	0.47	&	56	&		\\
NLTT 20303 AB	&	2.81	&	7.44	&	89.4	&	NLTT 20303B may be a very tight binary.\tablenotemark{b}	\\
1RXS 1014 AB	&	1.2	&	2.02	&	271	&		\\
2M 1036 AB	&	1.7	&	0.96	&	160	&		\\
NLTT 32659 AB	&	8.4	&	2.87	&	89	&		\\
1RXS 2043 AB	&	1.007	&	1.48	&	217	&		\\
G273-191 AB	&	1.057	&	1.96	&	176	&		\\
\enddata					
\tablenotetext{a}{This target, also known as G 274-24, was observed by \cite{daem07}.}
\end{deluxetable}

		\begin{deluxetable}{lcccc}
		\tabletypesize{\scriptsize}
		\tablewidth{0pt}
		\tablecolumns{5}
		\tablecaption{Binaries Resolved with Keck and Subaru Adaptive Optics Imaging\label{tab:aobinaries}}
		\tablehead{
		       \colhead{}               &  \colhead{UT Date of Observation}   &  \colhead{}                              &      \colhead{}                     &    \colhead{}    \\
		        \colhead{Target}   & \colhead{(Y/M/D)}   &       \colhead{Instrument}      &     \colhead{Filter}  &    \colhead{$\Delta$mag}    
		        }   
		\startdata
		G 217-32  AB                & 2010/08/18    &    NIRC2  &         $H $  &    0.475   (0.011)   \\
		1RXS J001557.5--163659 AB   & 2011/08/20    &    NIRC2  &         $H $  &    0.06    (0.05)    \\
		G 132-50 AB\tablenotemark{a}                 & 2011/12/28    &    HiCIAO &         $K_s$  &    2.37     (0.16)     \\
		NLTT 6549 AB               & 2011/08/20    &    NIRC2  &         $H $  &    1.45     (0.09)     \\
		G 81-34 AB                  & 2011/01/27    &    HiCIAO &         $K_s$  &    0.57     (0.02)    \\
		GJ 2079 AB                  & 2011/03/25    &    NIRC2  &         $K_s$  &    1.88     (0.06)\tablenotemark{b}     \\
		GJ 3629 AB                 & 2011/03/25    &    NIRC2  &         $K_s$  &    2.875    (0.012)\tablenotemark{c}    \\
		GJ 3729 AB                 & 2011/06/21    &    NIRC2  &         $H $  &    0.65    (0.06)    \\
\enddata
		\tablecomments{Binary separations at the time of discovery range from 0$\farcs$04--0$\farcs$65.  Details about the observations, 
		including astrometry for the binaries, will be presented in Bowler et al. (in preparation). }
		\tablenotetext{a}{G 132-50 AB is a common proper-motion companion to our spectroscopic targets G 132-51 B (E \& W). Including G 132-51 A, makes this a quintuple system in the AB Dor Moving Group.}
		\tablenotetext{b}{The peak counts from the primary ($\sim$1.2$\times$10$^4$~DN pix$^{-1}$) are in a regime where the NIRC2
		pixel response is modestly nonlinear ($\sim$10\%).  Before measuring flux ratios we therefore corrected these frames using the response curve derived by S. Metchev (private communication).}
		\tablenotetext{c}{The age ($\sim$25--300~Myr) and distance ($\sim$22~pc) of this system implies that GJ~3629 B 
		is a brown dwarf based on the evolutionary models of \cite{burr97}. The discovery of this substellar 
		companion is presented in Bowler et al. (2012, ApJ, submitted).}
		\tablenotetext{d}{Flux ratio measurement is from a single image.}
		\tablenotetext{e}{Details of this discovery are reported in \cite{bowl12a}.}
		\end{deluxetable}

\begin{deluxetable}{lcccccrrclccccc}							
\tabletypesize{\scriptsize}							
%\rotate							
\tablecaption{Radial Velocity Variable Targets \label{table_rv_variable}}							
\tablewidth{0pt}							
\tablehead{							
\colhead{VB Name}	&	\colhead{RV}			&	\colhead{Ref.}	\\
\colhead{}	&	\colhead{km~s${-1}$}			&	\colhead{}	
}							
\startdata							
							
II Tau	&	37.9	$\pm$	0.2	&	this work	\\
	&	43.1	$\pm$	0.6	&	this work	\\
	&	29.3	$\pm$	1.2	&	\cite{moch02}	\\
GJ 2079 (AB)	&	17.2	$\pm$	0.4	&	this work	\\
	&	10.0	$\pm$	1.0	&	\cite{mont01}	\\
	&	2.7	$\pm$	0.1	&	\cite{lope10}	\\
	&	4.9	$\pm$	1.5	&	\cite{gizi02b}	\\
GJ 9652 A 	&	-88.8	$\pm$	1.1	&	this work	\\
	&	-73.0	$\pm$	0.8	&	this work	\\
1RXS J193528.9+374605	&	-23.2	$\pm$	0.2	&	this work	\\
	&	-43.4	$\pm$	0.3	&	this work	\\

\enddata							
							
\end{deluxetable}

\clearpage

\end{document}